\documentclass[aps,prl,preprintnumbers,nofootinbib,amsmath,amssymb,superscriptaddress,10pt,twocolumn]{revtex4-2}

\usepackage{hyperref}
\usepackage{orcidlink}

\usepackage[nohyperlinks, nolist]{acronym}
\usepackage{graphicx}
\usepackage{xcolor}
\usepackage{multirow}
\usepackage{tabularx}
\usepackage{siunitx}
\usepackage{ulem}
\graphicspath{{./}{figures/}}

\newcommand{\red}[1]{\textcolor{black}{#1}}
\newcommand{\RED}[1]{\textcolor{black}{#1}}

\newcommand{\cit}[1]{{\small[{\textcolor{teal}{\textit{citation needed}}}}]}

\begin{document}


\title{Impact of the Einstein Telescope's duty cycle on the estimation of binary black hole parameters}
\newcommand{\UU}{Institute for Gravitational and Subatomic Physics (GRASP), Utrecht University, Princetonplein 1, 3584 CC Utrecht, The Netherlands}
\newcommand{\Nikhef}{Nikhef, Science Park 105, 1098 XG Amsterdam, The Netherlands}
\newcommand{\OU}{Faculty of Science, Open Universiteit, Valkenburgerweg 177, 6419 AT Heerlen, The Netherlands}
\newcommand{\UCL}{Centre for Cosmology, Particle Physics and Phenomenology - CP3, Universit\'e Catholique de Louvain, Louvain-La-Neuve, B-1348, Belgium}
\newcommand{\ORB}{Royal Observatory of Belgium, Avenue Circulaire, 3, 1180 Uccle, Belgium}
\newcommand{\KUL}{Leuven Gravity Institute, KU Leuven, Celestijnenlaan 200D box 2415, 3001 Leuven, Belgium}
\newcommand{\PHYKUL}{Department of Physics and Astronomy, Laboratory for Semiconductor Physics, KU Leuven, B-3001 Leuven, Belgium}
\newcommand{\EEKUL}{KU Leuven, Department of Electrical Engineering (ESAT), STADIUS Center for Dynamical Systems, Signal Processing and Data Analytics, B-3001 Leuven, Belgium}
\newcommand{\Upisa}{Dipartimento di Fisica "E. Fermi", Universit\`{a} di Pisa, I-56127 Pisa, Italy}
\newcommand{\UGent}{Department of Physics \& Astronomy, Ghent University, Proeftuinstraat 86, 9000 Ghent, Belgium}
\newcommand{\RWTH}{Physics Institute IIIB, RWTH Aachen University, Sommerfeldstr. 16, 52074 Aachen, Germany}
\newcommand{\CUHK}{Department of Physics, Chinese University of Hong Kong, Sha Tin, Hong Kong}

\author{Luca~Negri\,\orcidlink{0009-0001-5468-0721}}
\email{l.negri@uu.nl}
\affiliation{\UU}
\affiliation{\Nikhef}

\author{Thomas~C.~K.~Ng\,\orcidlink{0000-0002-9491-1598}}
\affiliation{\Nikhef}
\affiliation{\UU}

\author{Thibeau~Wouters\,\orcidlink{0009-0006-2797-3808}}
\affiliation{\UU}
\affiliation{\Nikhef}

\author{Tim~J.~Kuhlbusch\,\orcidlink{0000-0001-5699-2377}}
\affiliation{\RWTH}
\date{\today}

\author{Harsh~Narola\,\orcidlink{0000-0001-9161-7919}}
\affiliation{\UU}
\affiliation{\Nikhef}

\author{Robin~Chan\,\orcidlink{0009-0004-1594-7501}}
\affiliation{\ORB}
\affiliation{\UGent}

\author{Kailib~Ryan~Doney\,\orcidlink{0009-0000-8000-6231}}
\affiliation{\CUHK}
\affiliation{\KUL}

\author{Francesco~Cireddu\,\orcidlink{0009-0002-7074-4278}}
\affiliation{\KUL}
\affiliation{\PHYKUL}

\author{Isaac~C.~F.~Wong\,\orcidlink{0000-0003-2166-0027}}
\affiliation{\KUL}
\affiliation{\EEKUL}

\author{Fabian~Gittins\,\orcidlink{0000-0002-9439-7701}}
\affiliation{\UU}
\affiliation{\Nikhef}

\author{Peter~T.~H.~Pang\,\orcidlink{0000-0001-7041-3239}}
\affiliation{\Nikhef}
\affiliation{\UU}

\author{Anuradha~Samajdar\,\orcidlink{0000-0002-0857-6018}}
\affiliation{\UU}
\affiliation{\Nikhef}

\author{Achim~Stahl\,\orcidlink{0000-0002-8369-7506}}
\affiliation{\RWTH}

\author{Justin~Janquart\,\orcidlink{0000-0003-2888-7152}}
\affiliation{\ORB}
\affiliation{\UCL}

\author{Chris~Van~Den~Broeck\,\orcidlink{0000-0001-6800-4006}}
\affiliation{\UU}
\affiliation{\Nikhef}

\author{Tjonnie~G.~F.~Li\,\orcidlink{0000-0003-4297-7365}}
\affiliation{\KUL}
\affiliation{\PHYKUL}
\affiliation{\EEKUL}

\date{\today}

\begin{abstract}

The geometry of the Einstein Telescope, the proposed next-generation European gravitational-wave observatory, is yet to be finalized. Two competing designs are under consideration: a nested triangular configuration (ET-$\Delta$) and two separated L-shaped detectors (ET-2L).
Extensive prior comparisons of ET designs established the scientific landscape using the Fisher-information-matrix formalism and identified that duty-cycle-induced single-detector operation is precisely the regime where this approximation becomes less reliable, underscoring the need for a  \red{refined}, principled treatment of the duty cycle.  
In this manuscript, we build on that foundation by revisiting the comparison with full Bayesian parameter estimation of gravitational-wave signals from binary black-hole mergers, projected onto a simulated Einstein Telescope that incorporates a  \red{refined} duty cycle modelled via continuous-time Markov chains and testing different detector maintenance strategies.
We find that the redundancy inherent in the ET-$\Delta$ design enables it to maintain at least two operational  \red{detectors} for the majority of the observing time, whereas the ET-2L configuration is often limited to a single \red{detector}. Crucially, we show that\red{, during partial network operation, ET-$\Delta$ often outperforms ET-2L, and that} the increased multi-detector uptime translates into tighter constraints on the luminosity distance and source-frame component masses.
Notably, this remains true even when gravitational-wave events have a lower signal-to-noise ratio in ET-$\Delta$ than in ET-2L.

\end{abstract}

\maketitle

\acrodef{JSD}[JSD]{Jensen-Shannon divergence}
\acrodef{KL}[KL]{Kullback-Leibler divergence}
\acrodef{PE}[PE]{parameter estimation}
\acrodef{FIM}[FIM]{Fisher information matrix}
\acrodef{EMR}[EMR]{Euroregion Meuse-Rhine}
\acrodef{ET}[ET]{Einstein Telescope}
\acrodef{CE}[CE]{Cosmic Explorer}
\acrodef{LF}[LF]{low-frequency}
\acrodef{HF}[HF]{high-frequency}
\acrodef{MCMC}[MCMC]{Markov chain Monte Carlo}
\acrodef{GW}[GW]{gravitational wave}
\acrodef{EM}[EM]{electromagnetic}
\acrodef{LVK}[LVK]{LIGO-Virgo-KAGRA}
\acrodef{CBC}[CBC]{compact binary coalescence}
\acrodef{PN}[PN]{post-Newtonian}
\acrodef{BH}[BH]{black hole}
\acrodef{NS}[NS]{neutron star}
\acrodef{PSR}[PSR]{pulsar}
\acrodef{BBH}[BBH]{binary black hole}
\acrodef{BNS}[BNS]{binary neutron star}
\acrodef{NSBH}[NSBH]{neutron star-black hole}
\acrodef{EFT}[EFT]{effective field theory}
\acrodef{SNR}[SNR]{signal-to-noise ratio}
\acrodef{PSD}[PSD]{power spectral density}
\acrodef{3G}[3G]{third-generation}
\acrodef{CTMC}[CTMC]{continuous-time Markov chain}
\acrodef{SGWB}[SGWB]{stochastic gravitational wave background}

\section{Introduction}

Since the first direct detection of \acp{GW} in 2015~\cite{LIGOScientific:2016aoc, LIGOScientific:2016vlm}, observations from \acp{CBC} involving \acp{BH} and \acp{NS} have significantly expanded our understanding of our Universe. These signals enable tests of general relativity in the strong-field regime~\cite{LIGOScientific:2019fpa,LIGOScientific:2021sio,LIGOScientific:2020tif,LIGOScientific:2025wao,LIGOScientific:2026qni}; provide constraints on the properties of supranuclear matter~\cite{LIGOScientific:2018cki}; and are cosmic probes of the rapid expansion of the Universe~\cite{LIGOScientific:2017adf,LIGOScientific:2021aug}. With almost 400 detections to date by the \ac{LVK} Collaboration~\cite{LIGOScientific:2025pvj,LIGOScientific:2025slb,LIGOScientific:2025jau,LIGOScientific:2026wfs}, \ac{GW} astrophysics now provides unprecedented insight into cosmology and the population properties of compact objects. These efforts will be further bolstered by future ground-based observatories, including Europe's \ac{ET}~\cite{Hild:2008ng, Hild:2010id, Punturo:2010zz} and the United States' \ac{CE}~\cite{Reitze:2019iox, Evans:2021gyd, Evans:2023euw}, which are expected to observe $\mathcal{O}(10^5)$ events per year with substantially higher \acp{SNR}~\cite{ET:2019dnz,Kalogera:2021bya,Borhanian:2022czq,Branchesi:2023mws, ET:2025xjr}, providing the foundation for new discoveries.

To fully realize the scientific potential of these next-generation instruments, we must accurately estimate both the intrinsic parameters (e.g., mass and spin) and the extrinsic parameters (e.g., luminosity distance and sky location) of the signals' sources. While the accuracy of these measurements depends on the \ac{SNR}, the geometry of the \ac{GW} detector also plays a key role, especially in triangulating the source location. Among next-generation detectors, the design of \ac{ET} remains unsettled. There are two main proposals on the table. One proposal consists of three co-located, nested V-shaped  \red{detectors} with an opening angle of $\ang{60}$, commonly referred to as ET-$\Delta$. The other design includes two L-shaped detectors, either aligned with each other or misaligned by $\ang{45}$, located at separate sites in Europe, which we denote as ET-$2$L.

Numerous studies have been conducted to gauge the performance of these proposed designs~\cite{Hall:2019xmm,ET:2025xjr,Bhagwat:2023jwv,Branchesi:2023mws,Franciolini:2023opt,Gupta:2023lga,Maggiore:2024cwf,Iacovelli:2024mjy,Pedrotti:2025tfg,Begnoni:2025oyd,DeRenzis:2024dvx,Loffredo:2024gmx,Colombo:2025sdm,Begnoni:2025mtz,Nanadoumgar-Lacroze:2026lcw}. Building a complete picture of ET's science output requires going beyond two idealizations that these studies have adopted: a simplified treatment of the duty cycle, and the \ac{FIM} approximation for parameter estimation.

The duty cycle refers to the fraction of time an instrument operates during an observing run. We know from the current generation of \ac{LVK} interferometers that there are periods when a detector is not operational due to either maintenance or uncontrollable environmental factors. Indeed, the \ac{LVK} instruments have a fractional uptime of 60-80\% \cite{PhysRevD.111.062002}. While it is very hard to make predictions about the exact value before the instrumental design is finalized, a similar level of stability can be expected for \ac{ET}. 

Studies have also used the \ac{FIM} as a proxy for parameter estimation. While the \ac{FIM} facilitates fast predictions of parameter-estimation accuracy, it is well known to have several shortcomings~\cite{Vallisneri:2007ev,Rodriguez:2013mla,Dhani:2024jja}. Crucially, the accuracy and stability of the estimates degrade when fewer detectors are considered~\cite{Iacovelli:2022mbg,Dupletsa:2024gfl}, often with predictions for single detectors having to be discarded~\cite{Branchesi:2023mws}. Furthermore, the \ac{FIM} is unable to capture multimodality in posterior distributions~\cite{Santoliquido:2025aiq}, leading to overconfident and unrepresentative estimates of parameter recovery, as well as the inability to model complex correlation structures and distributions typical of some parameters, like the luminosity distance.

In this manuscript, we present a comprehensive comparison of the performance of the two proposed \ac{ET} designs. In contrast to previous works, the performance is quantified with both (i) a  \red{refined} treatment of the duty cycle, accounting for different assumptions and testing different maintenance strategies, and (ii) full parameter estimation on the simulated \ac{GW} signals.

\section{Methodology}\label{sec: methods}

\subsection{Duty cycle}\label{sec:methods-duty-cycle-estimates}
\ac{GW} detectors can become non-operational for a variety of reasons, including loss of resonance in the optical cavities, misalignment due to seismic activity or instrumental failures ~\cite{PhysRevD.111.062002,Virgo:2022ysc}, as well as scheduled maintenance, calibration, and upgrades. \red{Previous design comparisons \cite{Branchesi:2023mws,ET:2025xjr} use a purely statistical model of the duty cycle, with a fixed single detector uptime probability of 85\%. One of the aims of this manuscript is to provide a more refined model of the duty cycle and study the effects of relaxing the assumption of the 85\% figure.}

\red{ET instrumental design will have significant differences with current observatories~\cite{ETDesign:2020}: each detector is going to be composed of an \ac{HF} and a \ac{LF} dedicated interferometer, mirror suspensions are going to be improved, control schemes are going to be adapted for higher laser power and cryogenic components. but the different instrument generations are} expected to face similar challenges \red{to maintain detector stability}. Indeed, the LIGO network consists of two detectors (LIGO Hanford Observatory and LIGO Livingston Observatory \cite{LIGOScientific:2014pky}) with the same instrumental design. As \ac{ET} will also consist of multiple detectors of identical design, 
LIGO data is used as a proxy to simulate \ac{ET} up and down-times \red{average duration}.

In particular, we consider data from the LIGO detectors during the first part of the fourth observing run (O4a)~\cite{O4aLIGO}. We describe the state of the detector as ``up'' when science-ready data are available and ``down'' otherwise. Due to scheduled downtime on Tuesdays and Wednesdays \cite{Virgo:2022ypn}, the corresponding data are excluded, and uptime periods of less than $4$ seconds are also excluded due to their negligible impact on the observational scheme. Using \red{the period of O4a data, where both LIGO detectors were the most stable~\cite{PhysRevD.111.062002},} from June 1 to November 20, 2023, the uptime and downtime durations of single detectors, not including maintenance time, are found to be well approximated by an exponential distribution, with the average uptime $\lambda_u=\qty{5.4}{h}$ and average downtime $\lambda_d=\qty{1.6}{h}$. 
Details can be found in the Supplemental Material. 

\red{By varying the ratio between exponential decay rates we are able to obtain different up and downtime percentages. This allows us to test different assumptions for possible ET duty cycles.}

\subsubsection{Continuous-time Markov chain}
Given the observed exponential behaviour, we model a system of $N$ independent and identical detectors as a continuous-time Markov chain with $N+1$ states $i \in \{0,\dots,N\}$, where $i$ enumerates the number of online detectors. The Markov chain is fully described by the rate matrix $Q_{i j}$, with elements describing the exit rate from state $i$ to state $j$ for $i\neq j$~\cite{liggett_continuous_2010}.
The diagonal components $Q_{ii}$ are equal to the negated total exit rates of the state $i$; i.e.,
$Q_{ii} = -\sum_{i\neq j} Q_{ij}$.

Since the instruments are independent and identical, $Q_{i j}$ is entirely characterized by the individual detector exit rates for the online (up) $q_u=1/\lambda_u$ and offline (down) $q_d=1/\lambda_d$ states. 

To obtain the stationary distribution $\pi_i$, one solves the condition $\sum_i \pi_i Q_{i j} = 0$ with the constraint $\sum_i \pi_i = 1$~\cite{liggett_continuous_2010}.
The fraction of time with $i$ detectors of $N$ being online is
\begin{equation}
    \pi_i(N) = \left( N \atop i \right) \frac{(q_d/q_u)^i}{\left(1+q_d/q_u\right)^N}.
\end{equation}

For a more detailed description of continuous-time Markov chains we refer the reader to \cite{liggett_continuous_2010}. This formalism is modified in order to account for possible correlated unlocks for co-located detectors. Details can be found in the Supplemental Material.\\

\subsubsection{Correlated unlocks}
While \red{many} unlocks stem from internal causes, \red{some} are driven by external environmental factors such as heightened seismic activity and adverse weather.
We expect these effects to induce correlated unlocks across co-located detectors. 
Strong, distant earthquakes dominate the contribution of heightened ground motion, as opposed to those that are weak but close-by, such as the ones observed during the Virgo O3 run \cite{Virgo:2022ypn}. Their seismic waves propagate globally, unlocking detectors for hours and affecting the entire network regardless of separation. Since these types would result in correlated unlocks for both the ET-$\Delta$ and ET-2L designs, with the same effect on the duty cycle, we exclude this case from our analysis.

Periods of adverse weather are known to be correlated with more frequent unlock events \red{\cite{PhysRevD.111.062002,Virgo:2022ypn,VIRGO:2026okv}}. This can include strong winds or heightened microseismic noise driven by sea activity, and since elevated winds and sea activity are often correlated with each other, it is quite hard to decouple the effects, even if some studies \cite{Virgo:2022ypn} seem to hint that the main culprit for these unlocks is strong winds, capable of shaking the buildings in which the instrument is housed. Given that \ac{ET} will be situated \qty{200}{m} underground, its susceptibility to these weather effects is uncertain. We therefore conservatively assume that simultaneous weather-induced unlock accounts for 2\% of all downtime.

\subsubsection{Scheduled downtime}
Other than unexpected downtime, there are also periods scheduled for maintenance, upgrades, and calibration, which can be arranged to minimize their impact on the operations. The currently implemented strategy is to synchronize this across detectors, thus reducing the amount of time with only a single instrument operating, and thus improving the search efficiency and parameter estimation results.
For the ET-2L configuration, the same strategy would likely be implemented due to its similar design to current-generation instruments. 

However, for ET-$\Delta$, it has been proposed to undergo a rotating maintenance, i.e., leaving two  \red{detector}s online while performing maintenance on the third one~\cite{Hild:2008ng,Freise:2008dk,ETDesign:2020}. Regular maintenance is not expected to affect the observing capabilities of nearby detectors: the closest mirrors of different detectors are around \qty{400}{m} apart in ET-$\Delta$~\cite{ETtaskforce:2025}, distance comparable with the distance between the interferometer plane and the surface, \red{the \qty{600}{m} no-fly-zone for Virgo  ~\cite{Fiori:2020arj,LogbookBoschi:2019} and the \qty{200}{m} no-vehicle limit around Virgo's central and end buildings \cite{LogbookFiori:2019}}. \red{The assumption that maintenance would not disturb nearby detectors' holds as long as the activities performed do not involve any major operation, such as excavations or involves of heavy machinery. The access tunnels of one detector also need to be designed to not interfere with the operations of the others~\cite{Amann:2022pyq}}. The rotating maintenance strategy takes advantage of the inherent redundancy of the ET-$\Delta$ design and so far has not yet been taken into account in works comparing the two geometries. Also, as is the case with ET-2L with misaligned arms, any 2 detectors of ET-$\Delta$ are also able to measure the polarization of an incoming \ac{GW}, making a network operating only with 2 V-shaped detectors \red{nearly} as performing as a full ET-$\Delta$.

\begin{figure*}[t]
    \centering
    \includegraphics[width=\textwidth]{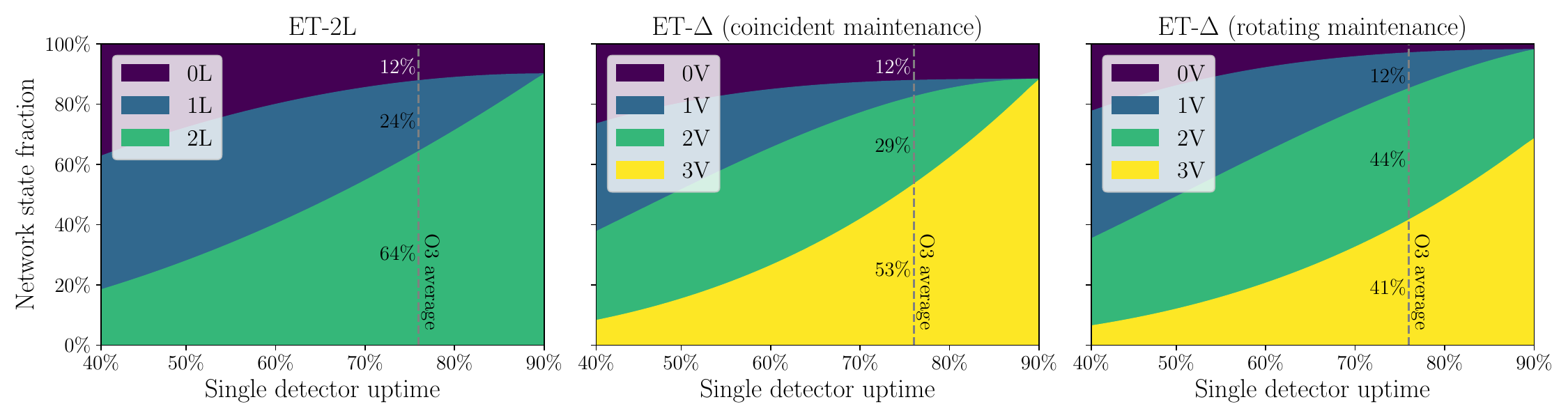}
    \caption{Fractions of time spent by ET-2L and ET-$\Delta$ in specific detector configurations as a function of the uptime of a single detector. Both coincident and rotating maintenance strategies are considered for ET-$\Delta$. Maintenance time is always set to 10\%. ET-$\Delta$ with rotating maintenance allows for both the highest chance of having at least 2 detectors online and the lowest chance of no detectors observing. Assuming single detector uptimes comparable with the O3 average of 76\% , these percentages are 85\% and 3\% respectively.}
    \label{fig:duty_cycle_estimate}
\end{figure*}

The network duty cycle percentages for the rest of this manuscript will report, for ET-$\Delta$, both the numbers obtained through the coincident and the rotating maintenance strategies, assuming that planned maintenance accounts for 10\% of the total downtime, roughly comparable to the values reported for current generation detectors \cite{Virgo:2022ypn,PhysRevD.111.062002,VIRGO:2026okv}.\red{This downtime represents the time dedicated to regular maintenance, calibration and minor upgrades that might happen during an observing run. This does not include the time that would be dedicated to major upgrades that usually take place between observing runs.} The resulting duty cycle estimates for ET-$\Delta$ and ET-2L can be seen in Fig. \ref{fig:duty_cycle_estimate}.

\subsection{Parameter estimation and simulation}\label{sec: methods PE}


In this work, we inject and analyze \RED{131} \ac{BBH} signals with parameters sampled randomly from the astrophisically motivated population presented in Ref. \cite{Regimbau:2012ir}. These injections were obtained by sampling 150 sources out of this population and selecting the ones that had a minimum SNR of 12 for the full \RED{ET-2L}, \RED{representing a population of signals which can be confidently detected by at least one full configuration}. These injections cover wide parameter ranges, with detector frame chirp masses ranging from 11 $M_\odot$ to 258 $M_\odot$ and redshifts from 0.5 to \RED{9}. The SNRs as measured by the full ET-$\Delta$ configuration go up to 180. We analyze each signal with the methods described in Refs.~\cite{Veitch:2009hd,Veitch:2014wba}. We analyze each signal in 5 different configurations: (i) the full ET-2L, (ii) ET-2L with one detector offline (always ET Sardinia), (iii) the full ET-$\Delta$, (iv) ET-$\Delta$ with one detector offline, and (v) ET-$\Delta$ with two detectors offline. The choice of which detector is offline does not bias the results, so the same detectors are always chosen. We simulate and analyze \ac{GW} signals using the \texttt{IMRPhenomXHM} waveform model of binary black-hole mergers~\cite{Garcia-Quiros:2020qpx}. 

The model includes higher-order modes, which can break the degeneracies between the luminosity distance and inclination angle~\cite{LIGOScientific:2020stg,Graff:2015bba,Kalaghatgi:2019log,Fairhurst:2023beb}. 
This waveform model assumes aligned black hole spins. We plan on studying the characterization of spin precession in a future extension of this work. \\

We place the simulated ET-$\Delta$ configuration with \qty{10}{km} arms at the site of the Virgo detector. For the ET-2L network, one  \red{detector} is located in the \ac{EMR} region, while the other is placed at the Sos Enattos site in Sardinia, each with \qty{15}{km} arms and a  $\ang{45}$ misalignment between the two detectors. 
 We simulate stationary Gaussian noise in the detectors using the \ac{PSD} curves from Ref.~\cite{Branchesi:2023mws}. Correlated noise is neglected for the co-located ET-$\Delta$ detectors~\cite{Janssens:2024jln}. The impact on parameter recovery depends on the sign and strength of the correlation \cite{Wong:2024hes}, and at the moment, it is not known which degree of correlation ET-$\Delta$ will experience. 
 We leave it to future work to include correlated noise in the likelihood, using the procedures described by, e.g., Refs. ~\cite{Cireddu:2023ssf,Liu:2024jna,Wong:2024hes,Caporali:2025mum}. \
The posterior distributions on the signal parameters $\theta$ are obtained with the \textsc{dynesty} nested sampler~\cite{Speagle:2019ivv,sergey_koposov_2024_12537467}, as implemented in the \textsc{bilby} inference package~\cite{Ashton:2018jfp,Romero-Shaw:2020owr}, with $2048$ live points to ensure robust recovery of the posterior.
To enhance the sampling efficiency, the luminosity distance is semi-analytically marginalized~\cite{Veitch:2014wba,farr2014_time_phase_marginalization,Singer:2015ema} and reconstructed afterward. The component masses measured in the detector frame $m^{\rm{det}}$  are affected by the redshift of the source $z$, through~\cite{Thrane:2018qnx}:

\begin{equation}
    m^{\mathrm{det}} = (1+z)m^{\rm{source}}
\end{equation}

Source masses $m^{\rm{source}}$ can be reconstructed if one has a measurement of $z$, which can be obtained through the luminosity distance by fixing a cosmological model. By default \textsc{bilby} uses the cosmology described in Ref.~\cite{Planck:2015fie}. The starting frequency at which the signals enter the detector is set to \qty{5}{\hertz}, with the duration of the analysis calculated accordingly. Since no signal in our dataset exceeds 256 seconds of duration we do not include the effects of earth's rotation.
For signals of duration longer than \qty{64}{s}, we use the relative-binning method to accelerate the likelihood evaluation~\cite{Cornish:2010kf,Zackay:2018qdy,Krishna:2023bug} and verify the results by computing the sample efficiency against the full likelihood after sampling \cite{Kong1992}. Out of the \RED{131} injections, \RED{34} used relative binning, and 3 of them \red{showed sampling efficiency below 1\% for at least 1 configuration. To avoid biasing results, these 3 signals were discarded completely from the analysis.}

\red{All of the posterior samples and the code to generate these results are available on Zenodo~\cite{negri_2026_20799710} } 

\section{Results}\label{sec: results}
\subsection{Duty cycle of the Einstein Telescope}\label{sec: results duty cycle estimates}

The estimated duty cycle for ET-2L and ET-$\Delta$ for different single detector uptimes is shown in Fig.~\ref{fig:duty_cycle_estimate}.
Across various operational capacities, the redundancy inherent to ET-$\Delta$, together with the rotational maintenance scheduling it enables, allows the observatory to operate with at least two detectors for a larger fraction of time compared to ET-2L, even when correlated downtime is considered.
 In particular, assuming single detector uptime to be comparable to the LVK average during O3 of 76\% \cite{PhysRevD.111.062002}, ET-$\Delta$ will operate with at least two  \red{detector}s for 85\% of the time for the rotating maintenance strategy and 82\% for the concurrent maintenance strategy, while ET-2L will have 2 interferometers online \RED{64}\% of the time. Using the rotating-maintenance scheme for ET-$\Delta$ significantly reduces the time without any detector to 3\%, at the cost of a shorter time with all detectors, from 53\% to 41\%.
 
\red{Our results differ from previous duty cycle estimates used in the literature. Using purely probabilistic models like the ones in \cite{Branchesi:2023mws,ET:2025xjr}, we find that by fixing the single detector uptime to 85\%, one find for ET-2L 72\%, 26\%, and 2\% for 2L, 1L, and 0L, respectively, while our model finds 80\%,9\% and 11\%. These differences are driven mainly by our different handling of the maintenance time, and therefore underestimate the percentage of time 2 detector are up at the same time. For ET-$\Delta$,  the probabilistic model finds 61\%,33\%,6\% and 0.3\%, while our model,  taking rotating maintenance as an example, finds 58\%, 36\%, 4\%, and 2\% for 3V,2V,1V and V, respectively. Rotating maintenance makes the two estimates quite similar, and the main differences are driven by other causes of correlated unlocks.}

\subsection{Parameter estimation}
\subsubsection{Special case: One detector offline}
\label{sec: results 2V vs 1L}
Our simulated GW signals are projected onto the detector operational configurations, i.e., ET-2L with one or two detectors operating, and ET-$\Delta$ with one, two, or three detectors operating. We first focus on a partial operational configuration of ET, comparing the parameter-estimation performance when one detector is offline, i.e., ET-2L operating with a single detector (ET-1L) and ET-$\Delta$ operating with two detectors (ET-2V).

\begin{figure}[htp]
    \centering
    \includegraphics[width=\columnwidth]{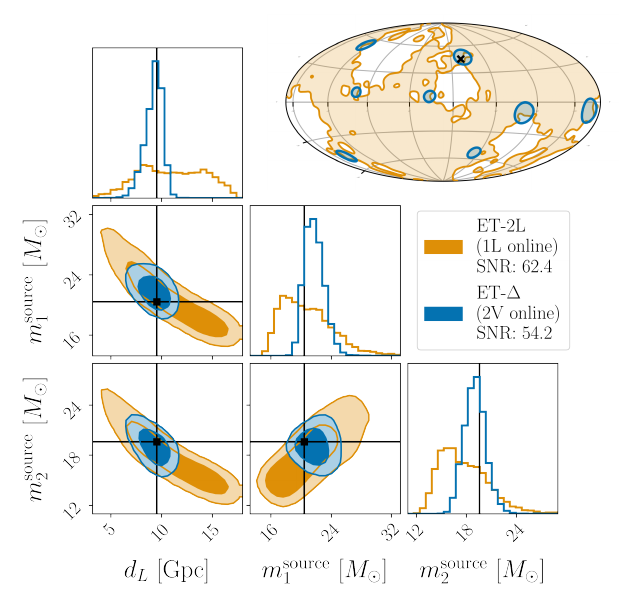}

    \caption{Posterior distributions of the source-frame component masses $m_1^\text{source}$ and $m_2^\text{source}$, the luminosity distance $d_{L}$ and the sky position, comparing 2V online (from the ET-$\Delta$ configuration) against 1L online (in the ET-2L design). 90\% credible intervals are drawn. Although the 1L online configuration yields a higher SNR than the 2V online configuration, the most astrophysically relevant parameters are better constrained by the 2V due to an improved measurement of the luminosity distance.  
    }
    \label{fig: example single injection 2V vs 1L}
\end{figure}

Figure~\ref{fig: example single injection 2V vs 1L} shows the posteriors of source-frame masses, luminosity distance, detector-frame chirp mass, and sky location for a single signal detected by both ET-$\Delta$ and ET-2L, each with one detector offline. The full posterior can be found in 
the Supplemental Material. The network \ac{SNR} of the signal in ET-1L is 62.4, which is higher than that of ET-2V, which is 54.2, leading to a better recovery of the detector-frame chirp mass and black hole spins for ET-1L 
However, the same does not hold for the extrinsic parameters, i.e., the luminosity distance and sky location, showcasing the non-trivial impact of detector geometry on parameter estimation. Moreover, since the measurement of the source frame masses is dependent on the recovered redshift, which in turn is obtained from the luminosity distance measurement by fixing a certain cosmological model, the posteriors on the source-frame masses are considerably tighter for ET-2V, further indicating that the geometry of the detector network can have a significant impact on science extraction.

\begin{figure}[htp]
    \centering
    \includegraphics[width=\columnwidth]{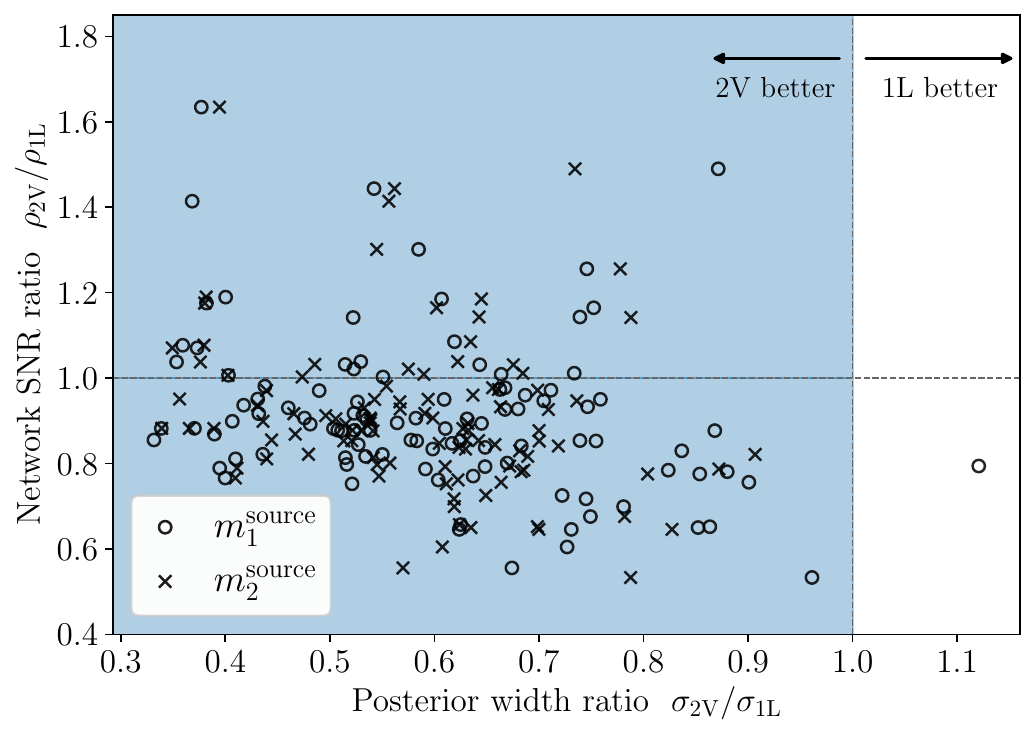}
    \caption{Ratio of the $90\%$ highest density interval widths of the posteriors for source-frame masses plotted against SNR ratio between ET-$\Delta$ with $2$V online and ET-2L with $1$L online. Independent of SNR ratio, ET-2V outperforms ET-1L in 99.6\% of instances. ET-2V posteriors are on average 0.6 times as wide as ET-1L. 
    Even if ET-1L imposes better constraints on the detector frame masses, this advantage is lost when looking at source frame masses, due to the improved accuracy in measuring redshift for ET-2V. 
    }
    \label{fig: 2V vs 1L combined MDC results}
\end{figure}

As shown in Fig.~\ref{fig: 2V vs 1L combined MDC results}, this trend is observed across the simulated population of signals. The figure shows a scatterplot of the ratio of the widths of the 90\% credible intervals for the source-frame component masses against the ratio of the network \ac{SNR} in the two ET configurations. While the majority of signals have a lower \ac{SNR} in ET-2V compared to ET-1L, the constraints on the source-frame masses from ET-2V are consistently tighter. With one detector offline in both configurations, ET-$\Delta$ yields better constraints on the source-frame masses than ET-2L, due to a more accurate estimate of the luminosity distance.

The improved constraints shown in Fig.~\ref{fig: example single injection 2V vs 1L} and Fig.~\ref{fig: 2V vs 1L combined MDC results} can be explained by the redundancy inherent to the triangular ET-$\Delta$ design. A single  \red{detector} can only resolve one polarization of the \ac{GW} signal, whereas the ET-$\Delta$ network (with one detector offline) still retains access to independent polarization content, reducing the sky-location posterior to eight well-localized modes~\cite{Singh:2020wsy, Santoliquido:2025lot} and in turn improving luminosity-distance measurements. For ET-2L, on the other hand, losing one detector leaves the network without polarization information, resulting in poor constraints on extrinsic parameters. Accurate measurements of the distance and source-frame masses can therefore only occur when both detectors in ET-2L are observing~\cite{Fairhurst:2009tc, KAGRA:2013rdx}.

Other comparisons between partial configurations are discussed in the Supplemental Material. 

\subsubsection{Simulating an observing scenario}
The above results demonstrate a few advantages of ET-$\Delta$ over ET-2L, specifically when one detector is offline. However, this is only one of multiple possible configurations. To obtain a realistic estimate of the performance of the two detector designs, one should repeat the comparison for each configuration and weight each result by the fraction of time the detector spends in that state.

To quantify the degradation of performance as a function of the detector uptimes, we compare the average performance of the instrument network with a fixed baseline configuration. Due to the longer arm length, the same signal detected by ET-2L with both detectors online will have the highest \ac{SNR}. Although this does not guarantee the most accurate posterior, we take this configuration as the reference. To quantify the degree of performance degradation as a function of the detector uptime, we define the efficiency $\epsilon$ as
\begin{equation}
    \epsilon(p,u)  = \sum^{\rm Detector \ states}_{i}\frac{\sigma_{\rm 2L}(p)}{\sigma_i(p)}w_i(u),
    \label{eq:efficiency}
\end{equation}
where $\sigma_{\rm 2L}(p)$ is the width of the $90\%$ credible interval of the parameter $p$ when the signal is detected by ET-2L with both detectors online, $\sigma_i(p)$ is the corresponding width for detector state $i$ of a given ET design, and $w_i(u)$ is the fraction of time the detector spends in that state as a function of the single detector uptime fraction $u$. By definition, an ET-2L with no downtime (an idealized and unrealistic scenario) has an efficiency of $1$, while an efficiency of $0$ corresponds to a non-operational detector. Therefore, the efficiency metric defined by Eq.~\eqref{eq:efficiency} quantifies how narrow, on average, we expect the reference configuration posteriors to be for a certain value of the uptime and a certain detector configuration. An efficiency of 0.5 indicates, for example, that the posteriors are on average twice as wide as the reference configuration. Since the reference configuration has been taken to be the one that leads to narrower constraints on average in ideal conditions, the efficiency will quantify the degradation of performance with respect to this baseline as a number between 0 and 1. 
\red{The efficiency metric is independent of the detectability of signals by different designs, since only parameter estimation accuracy and duty cycle estimates define the efficiency. Signal search algorithm will perform differently for different designs: the higher average SNR of ET-2L will have larger horizion distance, while null-stram based searches \cite{Macquet:2024qim} and a higher uptime with at least 2 detectors online can allow ET-$\Delta$ to achieve higher sensitivity. A full comparison between the two designs will have to consider the detectability of signals for the two designs given a certain astrophysical population, but this is beyond the scope of this manuscript.}

\begin{figure*}[t]
    \centering
    \includegraphics[width=\textwidth]{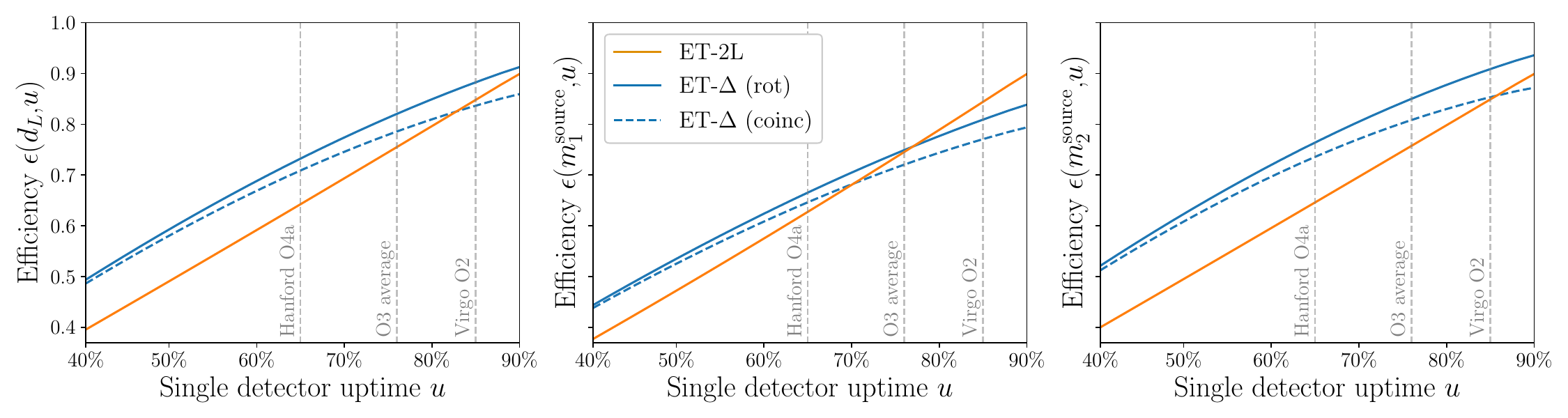}
    \caption{The efficiency of different \ac{ET} designs, defined by the duty-cycle-weighted uncertainty ratio (see Eq.~\ref{eq:efficiency}), as a function of the fractional uptime of a single detector, for luminosity distance and source-frame masses, both the rotational (rot) and coincident (coinc) maintenance schedules regarding for ET-$\Delta$. The uptimes of Hanford in O4a, Virgo in O2, and the overall O3 average offer a pessimistic, realistic, and optimistic projection, respectively. ET-$\Delta$ effciency is improved by implementing a rotating maintenance scheme, which was not explored in previous works. The configuration measures, on average, tighter confidence intervals than ET-2L for most of the parameter space explored
    , being outperformed only when ET-2L reaches average uptimes not yet achieved by any interferometer operational today.}
    \label{fig:money_plot}
\end{figure*}

Fig.~\ref{fig:money_plot} shows the median of the efficiency across the simulated population as a function of the fractional uptime of a single detector for the two ET designs. A higher value of the efficiency corresponds to an improved measurement of source parameters. A lower fractional single-detector uptime represents a less demanding and more achievable operational requirement. \red{The Virgo O2 uptime of around 85\% is similar to the value assumed from previous works that evaluate ET designs performances~\cite{Branchesi:2023mws,ET:2025xjr} } ET-$\Delta$ rotating maintenance outperforms the coincident maintenance in every scenario: the loss of time spent in a full $\Delta$ configuration is outweighed by the gain of time spent with at least one detector online. Considering this maintenance scheme, ET-$\Delta$ outperforms ET-2L across a wide range of detector capabilities, only showing a disadvantage at single-detector uptimes above 75\% for the primary source-frame mass and at least 90\% for the secondary source-frame mass and distance.\\

\subsubsection{Impact of a detector network}

When ET will be built, we expect it to be operational in conjunction with other third-generation  \red{detector}s, like \ac{CE} and LIGO India. Adding detectors to the network with a long baseline can greatly improve source localization~\cite{Iacovelli:2026ixl}, especially for the smaller detector networks, like ET-2L with one  \red{instrument} offline (ET-1L) and ET-$\Delta$ with two  \red{instruments}s offline (ET-1V). We repeated this study by adding a single \ac{CE}  \red{detector} with 40 km arm length positioned at the Hanford site. For the purpose of this study we do not simulate any duty cycle for CE and consider it always operational. We compute the efficiency of the 2 configuration taking the ET-2L + CE network working at 100\% uptime as a baseline, and we find that the crossover point between the efficiency of ET-2L and ET-$\Delta$ happens at 74\%, 80\%, and 78\% for $d_L$, $m_1^\text{source}$, and $m_2^\text{source}$, respectively. 
We plan to explore this in more detail in future works.

\section{Discussion}\label{sec: discussion}

\subsection{Limitations in duty cycle estimates}
This work presents, to the best of our knowledge, one of the most detailed studies of the impact of the duty cycle on \ac{ET}'s performance. However, several assumptions are made that warrant further investigation in future work.

The current \ac{ET} design proposes a xylophone configuration~\cite{Hild:2010id,Freise:2008dk}, in which each instrument comprises two interferometers, one optimized for the \ac{LF} and one for the \ac{HF}. In this work, we assume that the \ac{LF} and \ac{HF} interferometers within each instrument turn online or offline simultaneously. In practice, however, independent instrumental failures and differing environmental sensitivities of the two interferometer designs would break this correlation. \red{This partial operation of each detector due to the \ac{HF} and \ac{LF} interferometers indipendent unlocks may have different impact on the two ET designs. Some tentative exploration has already been conducted in~\cite{Branchesi:2023mws}, but a more in depth study is needed. Even tough}  we expect our qualitative conclusions to hold when the full duty-cycle landscape is considered, we \red{deem fundamental} to \red{test} this hypothesis in future works. 

The assumed fraction of correlated lock losses for co-located detectors also impacts our results. While current-generation detectors suggest this fraction is small, mainly due to adverse weather conditions, which will be of lesser impact to ET,  and major upgrades of the infrastructure, it remains difficult to predict how the control schemes, suspension designs, and seismic environment of \ac{ET} will combine to determine its duty cycle. Similarly, it is unclear whether future detectors will demand comparable maintenance to current facilities. The sensitivity of our results to these assumptions will be explored in future work.

\subsection{Other limitations}
This work uses a set of \RED{131} events sampled randomly from the population, representing the a typical observation of ET. This number is relatively low and it is only limited by the high computational costs of the runs. This could introduce statistical noise, and the dataset does not exhaustively cover all possible signals, excluding for example signals with very high SNRs, which will provide a lot of science output.

In this work, we do not take into account the specific advantages of the null stream, unique to ET-$\Delta$ ~\cite{Guersel:1989th,Freise:2008dk,Regimbau:2012ir}. We assume perfect knowledge of the \ac{PSD}, perfectly calibrated detectors, and absence of transient, non-Gaussian noise. The null stream could be used to improve constraints and reduce biases in a more realistic observing scenario, by providing an unbiased estimate of the detector \ac{PSD}~\cite{Regimbau:2012ir,Goncharov:2022dgl}, improving detector calibration~\cite{Schutz:2020hyz,Wong:2025iaf}, enabling vetoing of glitches and unbiased cleaning and removal ~\cite{Chatterji:2006nh,Narola:2024qdh}. We also do not include effects of possible correlated noise for co-located detectors, which can influence parameter estimation, especially in the lowest frequencies of ET~\cite{Cireddu:2023ssf,Janssens:2022xmo}. \red{Correlated noise is expected to come mainly form newtonian noise. Promising newtonian noise mitigation tecniques \cite{vanBeveren:2023seq} might help reduce the degree of correlations between co-located detectors}.

This work focuses only on \acp{BBH}, while ET will observe \acp{BNS} as well. Parameter estimation of \acp{BNS} in the ET era presents some advantages compared to \acp{BBH}: the long duration of these inspirals implies that the detection is affected by Earth rotation, which helps improve constraints for source localization ~\cite{Zhao:2017cbb}. The effect of the duty cycle on BNSs can be very complex: the long duration of a BNS in ET implies that instruments can come online or go offline during the event itself, and this can impact early-warning capabilities for the detection of electromagnetic counterparts. \red{Under the same assumptions, we do not expect the results showed in this manuscript to be substantially different for BNS signals}, but further investigations will be needed \red{to give a complete picture of the effects of duty cycle on BNSs}. 

\RED{Another source measurable by ET is the \ac{SGWB}~\cite{Caprini:2018mtu}, a constant signal formed by the superposition of many unresolvable sources, of either cosmological or astrophysical origin, detectable by correlating the antenna response of multiple detectors. Given detector uptimes of 100\%,  ET-$\Delta$ design outperforms the ET-2L design with misaligned arms \cite{Branchesi:2023mws,ET:2025xjr} but only if correlated noise for co-located detectors is correctly modeled and accounted for \cite{Caporali:2025mum}. By varying the duty cycle, since multiple online detectors are needed to measure the \ac{SGWB}, the higher multi-detector network uptimes for the ET-$\Delta$ should still make this the more sensitive design at all duty cycles; however, the characterization of correlated noise could be hindered by the absence of the null stream for the ET-2V configuration. If this is the case, a coincident maintenance scheme for ET-$\Delta$ would be preferred to maximize sensitivity to the \ac{SGWB}. Due to the uncertainties in the capabilities of modelling correlated noise, we cannot rule out the possibility of a reversed design hierarchy when it comes to detecting the \ac{SGWB} when accounting for the duty cycle. A failure to properly characterize and remove the \ac{SGWB} from the data could also impact and contaminate the parameter estimation of foreground sources, but we expect this effect to be negligible.}

\section{Conclusions}\label{sec: conclusion}
The geometry of the Einstein Telescope (ET) remains under discussion, with two proposed designs: a nested triangular detector (ET-$\Delta$) and two separated L-shaped detectors (ET-2L). Extensive studies to date, including Refs.~\cite{Branchesi:2023mws} and ~\cite{ET:2025xjr} established the scientific baseline using the Fisher-information-matrix formalism, and both noted that single-detector operation is where this approximation is least reliable. In this work, we have followed up on that observation by applying full Bayesian parameter estimation to a simulated population of \ac{BBH} mergers analyzed under a  \red{data-driven}, Markov-chain-derived duty cycle that also tests different planned-maintenance scenarios and accounts for correlated downtime among the co-located  \red{detector}s of ET-$\Delta$.

\red{Our model finds that, within the stated assumptions,} ET-$\Delta$ spends the majority of its observing time with at least two V-shaped \red{detector}s online, whereas for ET-2L, a good portion of the time is spent with only one detector online. Comparing estimation of parameters \red{of high scientific interest} \red{during partial network operation} 
we show that ET-$\Delta$ configurations with multiple detectors can outperform ET-2L configurations that achieve higher \acp{SNR}, demonstrating that the geometry of the detector network has a non-trivial impact on parameter estimation not captured by \ac{SNR} alone. These findings complement and refine the picture established by population-scale \ac{FIM} studies: the redundancy of ET-$\Delta$ confers advantages, particularly in luminosity distance and source-frame mass recovery, that are most apparent in the reduced-network configurations that dominate realistic observing scenarios, and that are difficult to capture with analytic approximations alone.

The increased multi-detector uptime of ET-$\Delta$ leads to consistently tighter constraints on the luminosity distance and source-frame component masses, which are key observables for population studies, cosmology, and gravitational lensing~\citep{Vijaykumar:2020pzn, Kumar:2021aog,Branchesi:2023mws}.  We find that for ET-$\Delta$, science output can be maximized by designing the instrument such that independent maintenance on the different detectors can be performed. This advantage holds across a wide range of detector capabilities.  \red{Regarding the parameter recovery, } ET-$\Delta$ is only being \red{definetly} outperformed at single-detector uptimes above 74\%, 
\red{making it a competitive design to }maximize ET's science output. Furthermore, the advantages reported here have not accounted for the additional benefits of the null stream in mitigating detector noise artifacts or estimating an unbiased PSD, ~\cite{Narola:2024qdh,Goncharov:2022dgl,Ebersold:2024hgp}, for example, further strengthening the case for the triangular design.\\

While further studies incorporating BNS sources, the xylophone configuration, \red{search algorithms effectiveness, extension to more science cases}, and the null stream advantages will refine this picture, our results already \red{show how in partial operation, the ET-$\Delta$ design can outperform ET-2L,} and \red{hint at an overall competitiveness}  for ET-$\Delta$ \red{when} realistic observing conditions \red{are taken into account}.

\section{Acknowledgments}\label{sec: acknowledgments}

We want to thank Jan Harms, Harald L{\"u}ck, and Tito dal Canton for their insightful comments provided during the internal review of this manuscript.   
We thank Sumit Kumar for valuable discussions regarding cosmology and population studies. 
We thank Filippo Santoliquido for help in running \texttt{ligo-skymap} to create skymaps and computing sky areas. 
We thank Jonathan Bratanata for the helpful discussions on the design of service tunnels for ET and their role in making rotating maintenance possible.

L.N. acknowledges support by the University of Utrecht (UU) and the Dutch National Institute for Subatomic Physics (Nikhef)

T.C.K.N. acknowledges support by the research program of the Netherlands Organization for Scientific Research (NWO).

T.W., H.N., and C.V.D.B. are supported by the research program of the Netherlands Organization for Scientific Research (NWO) through grant number OCENW.XL21.XL21.038. 

R.C. is supported by the Royal Observatory of Belgium, Fonds Wetenschappelĳk Onderzoek (FWO) research project G0A5E24N and FWO International Research Infrastructure (IRI) project I000725N.

F.G. acknowledges funding from the European Union’s Horizon Europe research and innovation programme under the Marie Sk{\l}odowska-Curie Grant Agreement No.~101151301.

P.T.H.P. is supported by the research program of the Netherlands Organization for Scientific Research (NWO) under grant number VI.Veni.232.021.

J.J. acknowledges support from the Fonds de la Recherche Scientifique (FNRS) and the Royal Observatory of Belgium.

T.G.F.L., I.C.F.W., F.C. are supported by the Research Foundation Flanders (FWO) research project G086722N and FWO International Research Infrastructure (IRI) projects I002123N and I000725N.

This work used the Dutch national e-infrastructure with the support of the
SURF Cooperative using grant no.~EINF-12633 and EINF-15686.

\bibliography{references}

@article{Narola:2024qdh,
    author = "Narola, Harsh and others",
    title = "{Null-stream-based third-generation-ready glitch mitigation for gravitational wave measurements}",
    eprint = "2411.15506",
    archivePrefix = "arXiv",
    primaryClass = "gr-qc",
    doi = "10.1103/l6tp-ykxp",
    journal = "Phys. Rev. D",
    volume = "112",
    number = "2",
    pages = "024079",
    year = "2025"
}

@article{Speagle:2019ivv,
    author = "Speagle, Joshua S.",
    title = "{dynesty: a dynamic nested sampling package for estimating Bayesian posteriors and evidences}",
    eprint = "1904.02180",
    archivePrefix = "arXiv",
    primaryClass = "astro-ph.IM",
    doi = "10.1093/mnras/staa278",
    journal = "Mon. Not. Roy. Astron. Soc.",
    volume = "493",
    number = "3",
    pages = "3132--3158",
    year = "2020"
}

@misc{sergey_koposov_2024_12537467,
  author       = {Sergey Koposov and
                  Josh Speagle and
                  Kyle Barbary and
                  Gregory Ashton and
                  Ed Bennett and
                  Johannes Buchner and
                  Carl Scheffler and
                  Ben Cook and
                  Colm Talbot and
                  James Guillochon and
                  Patricio Cubillos and
                  Andrés Asensio Ramos and
                  Matthieu Dartiailh and
                  Ilya and
                  Erik Tollerud and
                  Dustin Lang and
                  Ben Johnson and
                  jtmendel and
                  Edward Higson and
                  Thomas Vandal and
                  Tansu Daylan and
                  Ruth Angus and
                  patelR and
                  Phillip Cargile and
                  Patrick Sheehan and
                  Matt Pitkin and
                  Matthew Kirk and
                  Joel Leja and
                  joezuntz and
                  Danny Goldstein},
  title        = {joshspeagle/dynesty: v2.1.4},
  month        = jun,
  year         = 2024,
  publisher    = {Zenodo},
  version      = {v2.1.4},
  doi          = {10.5281/zenodo.12537467},
  url          = {https://doi.org/10.5281/zenodo.12537467},
}

@article{Ashton:2018jfp,
    author = "Ashton, Gregory and others",
    title = "{BILBY: A user-friendly Bayesian inference library for gravitational-wave astronomy}",
    eprint = "1811.02042",
    archivePrefix = "arXiv",
    primaryClass = "astro-ph.IM",
    doi = "10.3847/1538-4365/ab06fc",
    journal = "Astrophys. J. Suppl.",
    volume = "241",
    number = "2",
    pages = "27",
    year = "2019"
}

@article{Romero-Shaw:2020owr,
    author = "Romero-Shaw, I. M. and others",
    title = "{Bayesian inference for compact binary coalescences with bilby: validation and application to the first LIGO{\textendash}Virgo gravitational-wave transient catalogue}",
    eprint = "2006.00714",
    archivePrefix = "arXiv",
    primaryClass = "astro-ph.IM",
    doi = "10.1093/mnras/staa2850",
    journal = "Mon. Not. Roy. Astron. Soc.",
    volume = "499",
    number = "3",
    pages = "3295--3319",
    year = "2020"
}

@article{Garcia-Quiros:2020qpx,
    author = "Garc{\'\i}a-Quir{\'o}s, Cecilio and Colleoni, Marta and Husa, Sascha and Estell{\'e}s, H{\'e}ctor and Pratten, Geraint and Ramos-Buades, Antoni and Mateu-Lucena, Maite and Jaume, Rafel",
    title = "{Multimode frequency-domain model for the gravitational wave signal from nonprecessing black-hole binaries}",
    eprint = "2001.10914",
    archivePrefix = "arXiv",
    primaryClass = "gr-qc",
    doi = "10.1103/PhysRevD.102.064002",
    journal = "Phys. Rev. D",
    volume = "102",
    number = "6",
    pages = "064002",
    year = "2020"
}

@article{Zackay:2018qdy,
    author = "Zackay, Barak and Dai, Liang and Venumadhav, Tejaswi",
    title = "{Relative Binning and Fast Likelihood Evaluation for Gravitational Wave Parameter Estimation}",
    eprint = "1806.08792",
    archivePrefix = "arXiv",
    primaryClass = "astro-ph.IM",
    month = "6",
    year = "2018"
}

@article{Cornish:2010kf,
    author = "Cornish, Neil J.",
    title = "{Fast Fisher Matrices and Lazy Likelihoods}",
    eprint = "1007.4820",
    archivePrefix = "arXiv",
    primaryClass = "gr-qc",
    month = "7",
    year = "2010"
}

@article{Veitch:2014wba,
    author = "Veitch, J. and others",
    title = "{Parameter estimation for compact binaries with ground-based gravitational-wave observations using the LALInference software library}",
    eprint = "1409.7215",
    archivePrefix = "arXiv",
    primaryClass = "gr-qc",
    reportNumber = "LIGO-P1400152",
    doi = "10.1103/PhysRevD.91.042003",
    journal = "Phys. Rev. D",
    volume = "91",
    number = "4",
    pages = "042003",
    year = "2015"
}

@article{Veitch:2009hd,
    author = "Veitch, J. and Vecchio, A.",
    title = "{Bayesian coherent analysis of in-spiral gravitational wave signals with a detector network}",
    eprint = "0911.3820",
    archivePrefix = "arXiv",
    primaryClass = "astro-ph.CO",
    reportNumber = "LIGO-P0900117",
    doi = "10.1103/PhysRevD.81.062003",
    journal = "Phys. Rev. D",
    volume = "81",
    pages = "062003",
    year = "2010"
}

@techreport{farr2014_time_phase_marginalization,
  author       = {Farr, Will M.},
  title        = {Marginalisation of the Time and Phase Parameters in CBC Parameter Estimation},
  institution  = {LIGO Scientific Collaboration},
  year         = {2014},
  number       = {LIGO-T1400460},
  type         = {Technical Report},
  url          = {https://dcc.ligo.org/LIGO-T1400460/public},
  note         = {LIGO Document Control Center}
}

@article{Singer:2015ema,
    author = "Singer, Leo P. and Price, Larry R.",
    title = "{Rapid Bayesian position reconstruction for gravitational-wave transients}",
    eprint = "1508.03634",
    archivePrefix = "arXiv",
    primaryClass = "gr-qc",
    reportNumber = "LIGO-P1500009-V3, LIGO-P1500009-V4, LIGO-P1500009-V5, LIGO-P1500009-V6, LIGO-P1500009-V7, LIGO-P1500009-V8",
    doi = "10.1103/PhysRevD.93.024013",
    journal = "Phys. Rev. D",
    volume = "93",
    number = "2",
    pages = "024013",
    year = "2016"
}

@article{Branchesi:2023mws,
    author = "Branchesi, Marica and others",
    title = "{Science with the Einstein Telescope: a comparison of different designs}",
    eprint = "2303.15923",
    archivePrefix = "arXiv",
    primaryClass = "gr-qc",
    reportNumber = "ET-0084A-23",
    doi = "10.1088/1475-7516/2023/07/068",
    journal = "JCAP",
    volume = "07",
    pages = "068",
    year = "2023"
}

@article{Punturo:2010zz,
    author = "Punturo, M. and others",
    editor = "Ricci, Fulvio",
    title = "{The Einstein Telescope: A third-generation gravitational wave observatory}",
    doi = "10.1088/0264-9381/27/19/194002",
    journal = "Class. Quant. Grav.",
    volume = "27",
    pages = "194002",
    year = "2010"
}

@article{Hild:2010id,
    author = "Hild, S. and others",
    title = "{Sensitivity Studies for Third-Generation Gravitational Wave Observatories}",
    eprint = "1012.0908",
    archivePrefix = "arXiv",
    primaryClass = "gr-qc",
    doi = "10.1088/0264-9381/28/9/094013",
    journal = "Class. Quant. Grav.",
    volume = "28",
    pages = "094013",
    year = "2011"
}

@article{Hild:2008ng,
    author = "Hild, Stefan and Chelkowski, Simon and Freise, Andreas",
    title = "{Pushing towards the ET sensitivity using 'conventional' technology}",
    eprint = "0810.0604",
    archivePrefix = "arXiv",
    primaryClass = "gr-qc",
    month = "10",
    year = "2008"
}

@article{Reitze:2019iox,
    author = "Reitze, David and others",
    title = "{Cosmic Explorer: The U.S. Contribution to Gravitational-Wave Astronomy beyond LIGO}",
    eprint = "1907.04833",
    archivePrefix = "arXiv",
    primaryClass = "astro-ph.IM",
    reportNumber = "LIGO-P1900316",
    journal = "Bull. Am. Astron. Soc.",
    volume = "51",
    number = "7",
    pages = "035",
    year = "2019"
}

@article{Evans:2021gyd,
    author = "Evans, Matthew and others",
    title = "{A Horizon Study for Cosmic Explorer: Science, Observatories, and Community}",
    eprint = "2109.09882",
    archivePrefix = "arXiv",
    primaryClass = "astro-ph.IM",
    reportNumber = "CE-P2100003-v7, Cosmic Explorer technical report CE-P2100003-v6",
    month = "9",
    year = "2021"
}

@article{Evans:2023euw,
    author = "Evans, Matthew and others",
    title = "{Cosmic Explorer: A Submission to the NSF MPSAC ngGW Subcommittee}",
    eprint = "2306.13745",
    archivePrefix = "arXiv",
    primaryClass = "astro-ph.IM",
    month = "6",
    year = "2023"
}

@article{ET:2025xjr,
    author = "Abac, Adrian and others",
    collaboration = "ET",
    title = "{The Science of the Einstein Telescope}",
    eprint = "2503.12263",
    archivePrefix = "arXiv",
    primaryClass = "gr-qc",
    reportNumber = "ET-0036C-25",
    month = "3",
    year = "2025"
}

@article{ET:2019dnz,
    author = "Maggiore, Michele and others",
    collaboration = "ET",
    title = "{Science Case for the Einstein Telescope}",
    eprint = "1912.02622",
    archivePrefix = "arXiv",
    primaryClass = "astro-ph.CO",
    doi = "10.1088/1475-7516/2020/03/050",
    journal = "JCAP",
    volume = "03",
    pages = "050",
    year = "2020"
}

@article{Borhanian:2022czq,
    author = "Borhanian, Ssohrab and Sathyaprakash, B. S.",
    title = "{Listening to the Universe with next generation ground-based gravitational-wave detectors}",
    eprint = "2202.11048",
    archivePrefix = "arXiv",
    primaryClass = "gr-qc",
    doi = "10.1103/PhysRevD.110.083040",
    journal = "Phys. Rev. D",
    volume = "110",
    number = "8",
    pages = "083040",
    year = "2024"
}

@article{Kalogera:2021bya,
    author = "Kalogera, Vicky and others",
    title = "{The Next Generation Global Gravitational Wave Observatory: The Science Book}",
    eprint = "2111.06990",
    archivePrefix = "arXiv",
    primaryClass = "gr-qc",
    month = "11",
    year = "2021"
}

@book{liggett_continuous_2010,
	location = {Providence, {RI}},
	title = {},
	isbn = {978-0-8218-4949-1},
	series = {Graduate studies in mathematics 113},
	shorttitle = {Continuous time Markov processes},
	pagetotal = {xii+271},
	publisher = {American Math. Soc.},
	author = {Liggett, Thomas M.},
	year = {2010},
}

@misc{O4aLIGO,
  author = {{LIGO Collaboration}},
  title = {Timeline {O4a} {LIGO}},
  note = {Accessed 27th of September 2025},
  howpublished = "\url{https://gwosc.org/timeline/show/O4a_16KHZ_R1/H1_DATA*L1_DATA/1368195220/21260798/}",
}

@article{Gupta:2023lga,
    author = "Gupta, Ish and others",
    title = "{Characterizing gravitational wave detector networks: from A$^\sharp$ to cosmic explorer}",
    eprint = "2307.10421",
    archivePrefix = "arXiv",
    primaryClass = "gr-qc",
    reportNumber = "CE Document No. P2300019, CE Document No. P2300019-v2",
    doi = "10.1088/1361-6382/ad7b99",
    journal = "Class. Quant. Grav.",
    volume = "41",
    number = "24",
    pages = "245001",
    year = "2024"
}

@article{Santoliquido:2025aiq,
    author = "Santoliquido, Filippo and Tissino, Jacopo and Dupletsa, Ulyana and Branchesi, Marica and Harms, Jan",
    title = "{Comparing next-generation detector configurations for high-redshift gravitational wave sources with neural posterior estimation}",
    eprint = "2512.20699",
    archivePrefix = "arXiv",
    primaryClass = "gr-qc",
    month = "12",
    year = "2025"
}

@article{Santoliquido:2025lot,
    author = "Santoliquido, Filippo and others",
    title = "{Fast and accurate parameter estimation of high-redshift sources with the Einstein Telescope}",
    eprint = "2504.21087",
    archivePrefix = "arXiv",
    primaryClass = "astro-ph.HE",
    doi = "10.1103/wf1k-p5cl",
    journal = "Phys. Rev. D",
    volume = "112",
    number = "10",
    pages = "103015",
    year = "2025"
}

@article{Fairhurst:2009tc,
    author = "Fairhurst, Stephen",
    title = "{Triangulation of gravitational wave sources with a network of detectors}",
    eprint = "0908.2356",
    archivePrefix = "arXiv",
    primaryClass = "gr-qc",
    doi = "10.1088/1367-2630/11/12/123006",
    journal = "New J. Phys.",
    volume = "11",
    pages = "123006",
    year = "2009",
    note = "[Erratum: New J.Phys. 13, 069602 (2011)]"
}

@article{Vijaykumar:2020pzn,
    author = "Vijaykumar, Aditya and Saketh, M. V. S. and Kumar, Sumit and Ajith, Parameswaran and Choudhury, Tirthankar Roy",
    title = "{Probing the large scale structure using gravitational-wave observations of binary black holes}",
    eprint = "2005.01111",
    archivePrefix = "arXiv",
    primaryClass = "astro-ph.CO",
    doi = "10.1103/PhysRevD.108.103017",
    journal = "Phys. Rev. D",
    volume = "108",
    number = "10",
    pages = "103017",
    year = "2023"
}

@article{Kumar:2021aog,
    author = "Kumar, Sumit and Vijaykumar, Aditya and Nitz, Alexander H.",
    title = "{Detecting Baryon Acoustic Oscillations with Third-generation Gravitational Wave Observatories}",
    eprint = "2110.06152",
    archivePrefix = "arXiv",
    primaryClass = "astro-ph.CO",
    doi = "10.3847/1538-4357/ac5e34",
    journal = "Astrophys. J.",
    volume = "930",
    number = "2",
    pages = "113",
    year = "2022"
}

@article{LIGOScientific:2021aug,
    author = "Abbott, R. and others",
    collaboration = "LIGO Scientific, Virgo, KAGRA",
    title = "{Constraints on the Cosmic Expansion History from GWTC{\textendash}3}",
    eprint = "2111.03604",
    archivePrefix = "arXiv",
    primaryClass = "astro-ph.CO",
    reportNumber = "LIGO-P2100185-v6, LIGO-P2100185-v5",
    doi = "10.3847/1538-4357/ac74bb",
    journal = "Astrophys. J.",
    volume = "949",
    number = "2",
    pages = "76",
    year = "2023"
}

@article{KAGRA:2013rdx,
    author = "Abbott, B. P. and others",
    collaboration = "KAGRA, LIGO Scientific, Virgo",
    title = "{Prospects for observing and localizing gravitational-wave transients with Advanced LIGO, Advanced Virgo and KAGRA}",
    eprint = "1304.0670",
    archivePrefix = "arXiv",
    primaryClass = "gr-qc",
    reportNumber = "LIGO-P1200087, VIR-0288A-12, JGW-P1808427",
    doi = "10.1007/s41114-020-00026-9",
    journal = "Living Rev. Rel.",
    volume = "19",
    pages = "1",
    year = "2016"
}

@article{LIGOScientific:2016aoc,
    author = "Abbott, B. P. and others",
    collaboration = "LIGO Scientific, Virgo",
    title = "{Observation of Gravitational Waves from a Binary Black Hole Merger}",
    eprint = "1602.03837",
    archivePrefix = "arXiv",
    primaryClass = "gr-qc",
    reportNumber = "LIGO-P150914",
    doi = "10.1103/PhysRevLett.116.061102",
    journal = "Phys. Rev. Lett.",
    volume = "116",
    number = "6",
    pages = "061102",
    year = "2016"
}

@article{LIGOScientific:2016vlm,
    author = "Abbott, B. P. and others",
    collaboration = "LIGO Scientific, Virgo",
    title = "{Properties of the Binary Black Hole Merger GW150914}",
    eprint = "1602.03840",
    archivePrefix = "arXiv",
    primaryClass = "gr-qc",
    reportNumber = "LIGO-P1500218",
    doi = "10.1103/PhysRevLett.116.241102",
    journal = "Phys. Rev. Lett.",
    volume = "116",
    number = "24",
    pages = "241102",
    year = "2016"
}

@article{LIGOScientific:2019fpa,
    author = "Abbott, B. P. and others",
    collaboration = "LIGO Scientific, Virgo",
    title = "{Tests of General Relativity with the Binary Black Hole Signals from the LIGO-Virgo Catalog GWTC-1}",
    eprint = "1903.04467",
    archivePrefix = "arXiv",
    primaryClass = "gr-qc",
    reportNumber = "LIGO-P1800316",
    doi = "10.1103/PhysRevD.100.104036",
    journal = "Phys. Rev. D",
    volume = "100",
    number = "10",
    pages = "104036",
    year = "2019"
}

@article{LIGOScientific:2021sio,
    author = "Abbott, R. and others",
    collaboration = "LIGO Scientific, VIRGO, KAGRA",
    title = "{Tests of General Relativity with GWTC-3}",
    eprint = "2112.06861",
    archivePrefix = "arXiv",
    primaryClass = "gr-qc",
    reportNumber = "LIGO-P2100275",
    doi = "10.1103/PhysRevD.112.084080",
    journal = "Phys. Rev. D",
    volume = "112",
    number = "8",
    pages = "084080",
    year = "2025"
}

@article{LIGOScientific:2020tif,
    author = "Abbott, R. and others",
    collaboration = "LIGO Scientific, Virgo",
    title = "{Tests of general relativity with binary black holes from the second LIGO-Virgo gravitational-wave transient catalog}",
    eprint = "2010.14529",
    archivePrefix = "arXiv",
    primaryClass = "gr-qc",
    reportNumber = "LIGO-P2000091",
    doi = "10.1103/PhysRevD.103.122002",
    journal = "Phys. Rev. D",
    volume = "103",
    number = "12",
    pages = "122002",
    year = "2021"
}

@article{LIGOScientific:2018cki,
    author = "Abbott, B. P. and others",
    collaboration = "LIGO Scientific, Virgo",
    title = "{GW170817: Measurements of neutron star radii and equation of state}",
    eprint = "1805.11581",
    archivePrefix = "arXiv",
    primaryClass = "gr-qc",
    reportNumber = "LIGO-P1800115",
    doi = "10.1103/PhysRevLett.121.161101",
    journal = "Phys. Rev. Lett.",
    volume = "121",
    number = "16",
    pages = "161101",
    year = "2018"
}

@article{LIGOScientific:2025wao,
    author = "Abac, A. G. and others",
    collaboration = "LIGO Scientific, Virgo, KAGRA",
    title = "{Black Hole Spectroscopy and Tests of General Relativity with GW250114}",
    eprint = "2509.08099",
    archivePrefix = "arXiv",
    primaryClass = "gr-qc",
    reportNumber = "LIGO P2500461",
    doi = "10.1103/6c61-fm1n",
    journal = "Phys. Rev. Lett.",
    volume = "136",
    number = "4",
    pages = "041403",
    year = "2026"
}

@article{LIGOScientific:2025pvj,
    author = "Abac, A. G. and others",
    collaboration = "LIGO Scientific, VIRGO, KAGRA",
    title = "{GWTC-4.0: Population Properties of Merging Compact Binaries}",
    eprint = "2508.18083",
    archivePrefix = "arXiv",
    primaryClass = "astro-ph.HE",
    reportNumber = "LIGO-P2400004",
    month = "8",
    year = "2025"
}

@article{LIGOScientific:2025slb,
    author = "Abac, A. G. and others",
    collaboration = "LIGO Scientific, VIRGO, KAGRA",
    title = "{GWTC-4.0: Updating the Gravitational-Wave Transient Catalog with Observations from the First Part of the Fourth LIGO-Virgo-KAGRA Observing Run}",
    eprint = "2508.18082",
    archivePrefix = "arXiv",
    primaryClass = "gr-qc",
    reportNumber = "LIGO-P2400386",
    month = "8",
    year = "2025"
}

@article{LIGOScientific:2025jau,
    author = "Abac, A. G. and others",
    collaboration = "LIGO Scientific, VIRGO, KAGRA",
    title = "{GWTC-4.0: Constraints on the Cosmic Expansion Rate and Modified Gravitational-wave Propagation}",
    eprint = "2509.04348",
    archivePrefix = "arXiv",
    primaryClass = "astro-ph.CO",
    reportNumber = "LIGO-P2400152",
    month = "9",
    year = "2025"
}

@article{LIGOScientific:2017adf,
    author = "Abbott, B. P. and others",
    collaboration = "LIGO Scientific, Virgo, 1M2H, Dark Energy Camera GW-E, DES, DLT40, Las Cumbres Observatory, VINROUGE, MASTER",
    title = "{A gravitational-wave standard siren measurement of the Hubble constant}",
    eprint = "1710.05835",
    archivePrefix = "arXiv",
    primaryClass = "astro-ph.CO",
    reportNumber = "LIGO-P1700296, FERMILAB-PUB-17-472-A-AE",
    doi = "10.1038/nature24471",
    journal = "Nature",
    volume = "551",
    number = "7678",
    pages = "85--88",
    year = "2017"
}

@article{Dupletsa:2024gfl,
    author = "Dupletsa, Ulyana and Harms, Jan and Ng, Ken K. Y. and Tissino, Jacopo and Santoliquido, Filippo and Cozzumbo, Andrea",
    title = "{Validating prior-informed Fisher-matrix analyses against GWTC data}",
    eprint = "2404.16103",
    archivePrefix = "arXiv",
    primaryClass = "gr-qc",
    doi = "10.1103/PhysRevD.111.024036",
    journal = "Phys. Rev. D",
    volume = "111",
    number = "2",
    pages = "024036",
    year = "2025"
}

@article{Iacovelli:2022mbg,
    author = "Iacovelli, Francesco and Mancarella, Michele and Foffa, Stefano and Maggiore, Michele",
    title = "{GWFAST: A Fisher Information Matrix Python Code for Third-generation Gravitational-wave Detectors}",
    eprint = "2207.06910",
    archivePrefix = "arXiv",
    primaryClass = "astro-ph.IM",
    doi = "10.3847/1538-4365/ac9129",
    journal = "Astrophys. J. Supp.",
    volume = "263",
    number = "1",
    pages = "2",
    year = "2022"
}

@article{LIGOScientific:2026qni,
    author = "Abac, A. G. and others",
    collaboration = "LIGO Scientific, VIRGO, KAGRA",
    title = "{GWTC-4.0: Tests of General Relativity. I. Overview and General Tests}",
    eprint = "2603.19019",
    archivePrefix = "arXiv",
    primaryClass = "gr-qc",
    reportNumber = "LIGO-P2500065",
    month = "3",
    year = "2026"
}

@article{Bhagwat:2023jwv,
    author = "Bhagwat, Swetha and Pacilio, Costantino and Pani, Paolo and Mapelli, Michela",
    title = "{Landscape of stellar-mass black-hole spectroscopy with third-generation gravitational-wave detectors}",
    eprint = "2304.02283",
    archivePrefix = "arXiv",
    primaryClass = "gr-qc",
    reportNumber = "ET-0106A-23",
    doi = "10.1103/PhysRevD.108.043019",
    journal = "Phys. Rev. D",
    volume = "108",
    number = "4",
    pages = "043019",
    year = "2023"
}

@article{Franciolini:2023opt,
    author = "Franciolini, Gabriele and Iacovelli, Francesco and Mancarella, Michele and Maggiore, Michele and Pani, Paolo and Riotto, Antonio",
    title = "{Searching for primordial black holes with the Einstein Telescope: Impact of design and systematics}",
    eprint = "2304.03160",
    archivePrefix = "arXiv",
    primaryClass = "gr-qc",
    doi = "10.1103/PhysRevD.108.043506",
    journal = "Phys. Rev. D",
    volume = "108",
    number = "4",
    pages = "043506",
    year = "2023"
}

@article{Maggiore:2024cwf,
    author = "Maggiore, Michele and Iacovelli, Francesco and Belgacem, Enis and Mancarella, Michele and Muttoni, Niccol{\`o}",
    title = "{Comparison of global networks of third-generation gravitational-wave detectors}",
    eprint = "2411.05754",
    archivePrefix = "arXiv",
    primaryClass = "gr-qc",
    doi = "10.1088/1361-6382/ae110b",
    journal = "Class. Quant. Grav.",
    volume = "42",
    number = "21",
    pages = "215004",
    year = "2025"
}

@article{Krishna:2023bug,
    author = "Krishna, Kruthi and Vijaykumar, Aditya and Ganguly, Apratim and Talbot, Colm and Biscoveanu, Sylvia and George, Richard N. and Williams, Natalie and Zimmerman, Aaron",
    title = "{Accelerated parameter estimation in Bilby with relative binning}",
    eprint = "2312.06009",
    archivePrefix = "arXiv",
    primaryClass = "gr-qc",
    month = "12",
    year = "2023"
}

@article{Iacovelli:2024mjy,
    author = "Iacovelli, Francesco and Belgacem, Enis and Maggiore, Michele and Mancarella, Michele and Muttoni, Niccol{\`o}",
    title = "{Combining underground and on-surface third-generation gravitational-wave interferometers}",
    eprint = "2408.14946",
    archivePrefix = "arXiv",
    primaryClass = "gr-qc",
    doi = "10.1088/1475-7516/2024/10/085",
    journal = "JCAP",
    volume = "10",
    pages = "085",
    year = "2024"
}

@article{Nanadoumgar-Lacroze:2026lcw,
    author = "Nanadoumgar-Lacroze, Dounia and Muttoni, Niccol{\`o} and Maggiore, Michele and Mancarella, Michele",
    title = "{Cosmology and modified GW propagation from the BNS mass function at third-generation detector networks}",
    eprint = "2603.19377",
    archivePrefix = "arXiv",
    primaryClass = "gr-qc",
    month = "3",
    year = "2026"
}

@article{Pedrotti:2025tfg,
    author = "Pedrotti, Alessandro and Mancarella, Michele and Bel, Julien and Gerosa, Davide",
    title = "{Cosmology with the angular cross-correlation of gravitational-wave and galaxy catalogs: forecasts for next-generation interferometers and the Euclid survey}",
    eprint = "2504.10482",
    archivePrefix = "arXiv",
    primaryClass = "astro-ph.CO",
    month = "4",
    year = "2025"
}

@article{Begnoni:2025oyd,
    author = "Begnoni, Andrea and Anselmi, Stefano and Pieroni, Mauro and Renzi, Alessandro and Ricciardone, Angelo",
    title = "{Detectability and Parameter Estimation for Einstein Telescope Configurations with GWJulia}",
    eprint = "2506.21530",
    archivePrefix = "arXiv",
    primaryClass = "astro-ph.CO",
    month = "6",
    year = "2025"
}

@article{DeRenzis:2024dvx,
    author = "De Renzis, Viola and Iacovelli, Francesco and Gerosa, Davide and Mancarella, Michele and Pacilio, Costantino",
    title = "{Forecasting the population properties of merging black holes}",
    eprint = "2410.17325",
    archivePrefix = "arXiv",
    primaryClass = "astro-ph.HE",
    doi = "10.1103/PhysRevD.111.044048",
    journal = "Phys. Rev. D",
    volume = "111",
    number = "4",
    pages = "044048",
    year = "2025"
}

@article{Caporali:2025mum,
    author = "Caporali, Ilaria and Capurri, Giulia and Del Pozzo, Walter and Ricciardone, Angelo and Valbusa Dall'Armi, Lorenzo",
    title = "{Impact of correlated noise on the reconstruction of the stochastic gravitational wave background with the Einstein Telescope}",
    eprint = "2501.09057",
    archivePrefix = "arXiv",
    primaryClass = "gr-qc",
    doi = "10.1103/s24t-k39c",
    journal = "Phys. Rev. D",
    volume = "112",
    number = "2",
    pages = "022005",
    year = "2025"
}

@article{Colombo:2025sdm,
    author = "Colombo, Alberto and Salafia, Om Sharan and Ghirlanda, Giancarlo and Iacovelli, Francesco and Mancarella, Michele and Broekgaarden, Floor S. and Nava, Lara and Giacomazzo, Bruno and Colpi, Monica",
    title = "{Multi-messenger observations in the Einstein Telescope era: Binary neutron star and black hole{\textendash}neutron star mergers}",
    eprint = "2503.00116",
    archivePrefix = "arXiv",
    primaryClass = "astro-ph.HE",
    doi = "10.1051/0004-6361/202554326",
    journal = "Astron. Astrophys.",
    volume = "704",
    pages = "A260",
    year = "2025"
}

@article{Ebersold:2024hgp,
    author = "Ebersold, Michael and Regimbau, Tania and Christensen, Nelson",
    title = "{Next-generation global gravitational-wave detector network: Impact of detector orientation on compact binary coalescence and stochastic gravitational-wave background searches}",
    eprint = "2408.06032",
    archivePrefix = "arXiv",
    primaryClass = "gr-qc",
    doi = "10.1103/PhysRevD.110.122006",
    journal = "Phys. Rev. D",
    volume = "110",
    number = "12",
    pages = "122006",
    year = "2024"
}

@article{Loffredo:2024gmx,
    author = "Loffredo, E. and others",
    title = "{Prospects for optical detections from binary neutron star mergers with the next-generation multi-messenger observatories}",
    eprint = "2411.02342",
    archivePrefix = "arXiv",
    primaryClass = "astro-ph.HE",
    doi = "10.1051/0004-6361/202452863",
    journal = "Astron. Astrophys.",
    volume = "697",
    pages = "A36",
    year = "2025"
}

@article{Begnoni:2025mtz,
    author = "Begnoni, Andrea and Del Pozzo, Walter and Pegorin, Matteo and Pomper, Joachim and Ricciardone, Angelo",
    title = "{Tests of General Relativity with Einstein Telescope}",
    eprint = "2511.07520",
    archivePrefix = "arXiv",
    primaryClass = "gr-qc",
    month = "11",
    year = "2025"
}

@article{Rodriguez:2013mla,
    author = "Rodriguez, Carl L. and Farr, Benjamin and Farr, Will M. and Mandel, Ilya",
    title = "{Inadequacies of the Fisher Information Matrix in gravitational-wave parameter estimation}",
    eprint = "1308.1397",
    archivePrefix = "arXiv",
    primaryClass = "astro-ph.IM",
    doi = "10.1103/PhysRevD.88.084013",
    journal = "Phys. Rev. D",
    volume = "88",
    number = "8",
    pages = "084013",
    year = "2013"
}

@article{Vallisneri:2007ev,
    author = "Vallisneri, Michele",
    title = "{Use and abuse of the Fisher information matrix in the assessment of gravitational-wave parameter-estimation prospects}",
    eprint = "gr-qc/0703086",
    archivePrefix = "arXiv",
    reportNumber = "LIGO-P070009-00-Z",
    doi = "10.1103/PhysRevD.77.042001",
    journal = "Phys. Rev. D",
    volume = "77",
    pages = "042001",
    year = "2008"
}

@article{Hall:2019xmm,
    author = "Hall, Evan D. and Evans, Matthew",
    title = "{Metrics for next-generation gravitational-wave detectors}",
    eprint = "1902.09485",
    archivePrefix = "arXiv",
    primaryClass = "astro-ph.IM",
    reportNumber = "LIGO-P1800304",
    doi = "10.1088/1361-6382/ab41d6",
    journal = "Class. Quant. Grav.",
    volume = "36",
    number = "22",
    pages = "225002",
    year = "2019"
}

@article{Dhani:2024jja,
    author = {Dhani, Arnab and V{\"o}lkel, Sebastian H. and Buonanno, Alessandra and Estelles, Hector and Gair, Jonathan and Pfeiffer, Harald P. and Pompili, Lorenzo and Toubiana, Alexandre},
    title = "{Systematic Biases in Estimating the Properties of Black Holes Due to Inaccurate Gravitational-Wave Models}",
    eprint = "2404.05811",
    archivePrefix = "arXiv",
    primaryClass = "gr-qc",
    doi = "10.1103/5pks-qz6b",
    journal = "Phys. Rev. X",
    volume = "15",
    number = "3",
    pages = "031036",
    year = "2025"
}

@article{Freise:2008dk,
    author = "Freise, A. and Chelkowski, S. and Hild, S. and Del Pozzo, W. and Perreca, A. and Vecchio, A.",
    title = "{Triple Michelson Interferometer for a Third-Generation Gravitational Wave Detector}",
    eprint = "0804.1036",
    archivePrefix = "arXiv",
    primaryClass = "gr-qc",
    doi = "10.1088/0264-9381/26/8/085012",
    journal = "Class. Quant. Grav.",
    volume = "26",
    pages = "085012",
    year = "2009"
}

@article{Fairhurst:2023beb,
    author = "Fairhurst, Stephen and Mills, Cameron and Colpi, Monica and Schneider, Raffaella and Sesana, Alberto and Trinca, Alessandro and Valiante, Rosa",
    title = "{Identifying heavy stellar black holes at cosmological distances with next-generation gravitational-wave observatories}",
    eprint = "2310.18158",
    archivePrefix = "arXiv",
    primaryClass = "astro-ph.HE",
    doi = "10.1093/mnras/stae443",
    journal = "Mon. Not. Roy. Astron. Soc.",
    volume = "529",
    number = "3",
    pages = "2116--2130",
    year = "2024"
}

@article{LIGOScientific:2020stg,
    author = "Abbott, R. and others",
    collaboration = "LIGO Scientific, Virgo",
    title = "{GW190412: Observation of a Binary-Black-Hole Coalescence with Asymmetric Masses}",
    eprint = "2004.08342",
    archivePrefix = "arXiv",
    primaryClass = "astro-ph.HE",
    reportNumber = "LIGO-P190412",
    doi = "10.1103/PhysRevD.102.043015",
    journal = "Phys. Rev. D",
    volume = "102",
    number = "4",
    pages = "043015",
    year = "2020"
}

@article{Kalaghatgi:2019log,
    author = "Kalaghatgi, Chinmay and Hannam, Mark and Raymond, Vivien",
    title = "{Parameter estimation with a spinning multimode waveform model}",
    eprint = "1909.10010",
    archivePrefix = "arXiv",
    primaryClass = "gr-qc",
    doi = "10.1103/PhysRevD.101.103004",
    journal = "Phys. Rev. D",
    volume = "101",
    number = "10",
    pages = "103004",
    year = "2020"
}

@article{Graff:2015bba,
    author = "Graff, Philip B. and Buonanno, Alessandra and Sathyaprakash, B. S.",
    title = "{Missing Link: Bayesian detection and measurement of intermediate-mass black-hole binaries}",
    eprint = "1504.04766",
    archivePrefix = "arXiv",
    primaryClass = "gr-qc",
    doi = "10.1103/PhysRevD.92.022002",
    journal = "Phys. Rev. D",
    volume = "92",
    number = "2",
    pages = "022002",
    year = "2015"
}

@article{Regimbau:2012ir,
    author = "Regimbau, Tania and others",
    title = "{A Mock Data Challenge for the Einstein Gravitational-Wave Telescope}",
    eprint = "1201.3563",
    archivePrefix = "arXiv",
    primaryClass = "gr-qc",
    doi = "10.1103/PhysRevD.86.122001",
    journal = "Phys. Rev. D",
    volume = "86",
    pages = "122001",
    year = "2012"
}

@article{Goncharov:2022dgl,
    author = "Goncharov, Boris and Nitz, Alexander H. and Harms, Jan",
    title = "{Utilizing the null stream of the Einstein Telescope}",
    eprint = "2204.08533",
    archivePrefix = "arXiv",
    primaryClass = "gr-qc",
    doi = "10.1103/PhysRevD.105.122007",
    journal = "Phys. Rev. D",
    volume = "105",
    number = "12",
    pages = "122007",
    year = "2022"
}

@article{Schutz:2020hyz,
    author = "Schutz, B. F. and Sathyaprakash, B. S.",
    title = "{Self-calibration of Networks of Gravitational Wave Detectors}",
    eprint = "2009.10212",
    archivePrefix = "arXiv",
    primaryClass = "gr-qc",
    month = "9",
    year = "2020"
}

@article{Guersel:1989th,
    author = "Guersel, Y. and Tinto, M.",
    title = "{Near optimal solution to the inverse problem for gravitational wave bursts}",
    doi = "10.1103/PhysRevD.40.3884",
    journal = "Phys. Rev. D",
    volume = "40",
    pages = "3884--3938",
    year = "1989"
}

@article{Wong:2025iaf,
    author = "Wong, Isaac C. F. and Cireddu, Francesco and Wils, Milan and Colemont, Tom and Narola, Harsh and Van Den Broeck, Chris and Li, Tjonnie G. F.",
    title = "{Bayesian Calibration of Gravitational-Wave Detectors Using Null Streams Without Waveform Assumptions}",
    eprint = "2510.06327",
    archivePrefix = "arXiv",
    primaryClass = "gr-qc",
    month = "10",
    year = "2025"
}

@article{Singh:2020wsy,
    author = "Singh, Neha and Bulik, Tomasz",
    title = "{Constraining parameters of coalescing stellar mass binary black hole systems with the Einstein Telescope alone}",
    eprint = "2011.06336",
    archivePrefix = "arXiv",
    primaryClass = "astro-ph.HE",
    reportNumber = "Virgo document number VIR-0876B-20",
    doi = "10.1103/PhysRevD.104.043014",
    journal = "Phys. Rev. D",
    volume = "104",
    number = "4",
    pages = "043014",
    year = "2021"
}

@article{Wong:2024hes,
    author = "Wong, Isaac C. F. and Pang, Peter T. H. and Wils, Milan and Cireddu, Francesco and Del Pozzo, Walter and Li, Tjonnie G. F.",
    title = "{Potential impact of noise correlation in next-generation gravitational wave detectors}",
    eprint = "2407.08728",
    archivePrefix = "arXiv",
    primaryClass = "gr-qc",
    doi = "10.1103/PhysRevD.111.044046",
    journal = "Phys. Rev. D",
    volume = "111",
    number = "4",
    pages = "044046",
    year = "2025"
}

@article{Cireddu:2023ssf,
    author = "Cireddu, Francesco and Wils, Milan and Wong, Isaac C. F. and Pang, Peter T. H. and Li, Tjonnie G. F. and Del Pozzo, Walter",
    title = "{Likelihood for a network of gravitational-wave detectors with correlated noise}",
    eprint = "2312.14614",
    archivePrefix = "arXiv",
    primaryClass = "gr-qc",
    doi = "10.1103/PhysRevD.110.104060",
    journal = "Phys. Rev. D",
    volume = "110",
    number = "10",
    pages = "104060",
    year = "2024"
}

@article{Liu:2024jna,
    author = "Liu, Jianan and Vajpeyi, Avi and Meyer, Renate and Janssens, Kamiel and Lee, Jeung Eun and Maturana-Russel, Patricio and Christensen, Nelson and Liu, Yixuan",
    title = "{Variational inference for correlated gravitational wave detector network noise}",
    eprint = "2409.13224",
    archivePrefix = "arXiv",
    primaryClass = "gr-qc",
    doi = "10.1103/PhysRevD.111.062003",
    journal = "Phys. Rev. D",
    volume = "111",
    number = "6",
    pages = "062003",
    year = "2025"
}

@article{Janssens:2024jln,
    author = "Janssens, Kamiel and others",
    title = "{Correlated 0.01{\textendash}40~Hz seismic and Newtonian noise and its impact on future gravitational-wave detectors}",
    eprint = "2402.17320",
    archivePrefix = "arXiv",
    primaryClass = "gr-qc",
    doi = "10.1103/PhysRevD.109.102002",
    journal = "Phys. Rev. D",
    volume = "109",
    number = "10",
    pages = "102002",
    year = "2024"
}

@article{Chatterji:2006nh,
    author = "Chatterji, Shourov and Lazzarini, Albert and Stein, Leo and Sutton, Patrick J. and Searle, Antony and Tinto, Massimo",
    title = "{Coherent network analysis technique for discriminating gravitational-wave bursts from instrumental noise}",
    eprint = "gr-qc/0605002",
    archivePrefix = "arXiv",
    reportNumber = "LIGO-P060009-01-E",
    doi = "10.1103/PhysRevD.74.082005",
    journal = "Phys. Rev. D",
    volume = "74",
    pages = "082005",
    year = "2006"
}

@article{Macquet:2024qim,
    author = "Macquet, Adrian and Dal Canton, Tito and Regimbau, Tania",
    title = "{Weakly modeled search for compact binary coalescences in the Einstein Telescope}",
    eprint = "2408.13007",
    archivePrefix = "arXiv",
    primaryClass = "gr-qc",
    doi = "10.1103/PhysRevD.111.022002",
    journal = "Phys. Rev. D",
    volume = "111",
    number = "2",
    pages = "022002",
    year = "2025"
}

@article{Zhao:2017cbb,
    author = "Zhao, Wen and Wen, Linqing",
    title = "{Localization accuracy of compact binary coalescences detected by the third-generation gravitational-wave detectors and implication for cosmology}",
    eprint = "1710.05325",
    archivePrefix = "arXiv",
    primaryClass = "astro-ph.CO",
    doi = "10.1103/PhysRevD.97.064031",
    journal = "Phys. Rev. D",
    volume = "97",
    number = "6",
    pages = "064031",
    year = "2018"
}

@article{PhysRevD.111.062002,
  title = {Advanced LIGO detector performance in the fourth observing run},
  author = "Capote, E. and Jia, W. and Aritomi, N. and Nakano, M. and Xu, V. and Abbott, R. and Abouelfettouh, I. and Adhikari, R. X. and Ananyeva, A. and Appert, S. and Apple, S. K. and Arai, K. and Aston, S. M. and Ball, M. and Ballmer, S. W. and Barker, D. and Barsotti, L. and Berger, B. K. and Betzwieser, J. and Bhattacharjee, D. and Billingsley, G. and Biscans, S. and Blair, C. D. and Bode, N. and Bonilla, E. and Bossilkov, V. and Branch, A. and Brooks, A. F. and Brown, D. D. and Bryant, J. and Cahillane, C. and Cao, H. and Clara, F. and Collins, J. and Compton, C. M. and Cottingham, R. and Coyne, D. C. and Crouch, R. and Csizmazia, J. and Cumming, A. and Dartez, L. P. and Davis, D. and Demos, N. and Dohmen, E. and Driggers, J. C. and Dwyer, S. E. and Effler, A. and Ejlli, A. and Etzel, T. and Evans, M. and Feicht, J. and Frey, R. and Frischhertz, W. and Fritschel, P. and Frolov, V. V. and Fuentes-Garcia, M. and Fulda, P. and Fyffe, M. and Ganapathy, D. and Gateley, B. and Gayer, T. and Giaime, J. A. and Giardina, K. D. and Glanzer, J. and Goetz, E. and Goetz, R. and Goodwin-Jones, A. W. and Gras, S. and Gray, C. and Griffith, D. and Grote, H. and Guidry, T. and Gurs, J. and Hall, E. D. and Hanks, J. and Hanson, J. and Heintze, M. C. and Helmling-Cornell, A. F. and Holland, N. A. and Hoyland, D. and Huang, H. Y. and Inoue, Y. and James, A. L. and Jamies, A. and Jennings, A. and Jones, D. H. and Kabagoz, H. B. and Karat, S. and Karki, S. and Kasprzack, M. and Kawabe, K. and Kijbunchoo, N. and King, P. J. and Kissel, J. S. and Komori, K. and Kontos, A. and Kumar, Rahul and Kuns, K. and Landry, M. and Lantz, B. and Laxen, M. and Lee, K. and Lesovsky, M. and Villarreal, F. Llamas and Lormand, M. and Loughlin, H. A. and Macas, R. and MacInnis, M. and Makarem, C. N. and Mannix, B. and Mansell, G. L. and Martin, R. M. and Mason, K. and Matichard, F. and Mavalvala, N. and Maxwell, N. and McCarrol, G. and McCarthy, R. and McClelland, D. E. and McCormick, S. and McRae, T. and Mera, F. and Merilh, E. L. and Meylahn, F. and Mittleman, R. and Moraru, D. and Moreno, G. and Mullavey, A. and Nelson, T. J. N. and Neunzert, A. and Notte, J. and Oberling, J. and O'Hanlon, T. and Osthelder, C. and Ottaway, D. J. and Overmier, H. and Parker, W. and Patane, O. and Pele, A. and Pham, H. and Pirello, M. and Pullin, J. and Quetschke, V. and Ramirez, K. E. and Ransom, K. and Reyes, J. and Richardson, J. W. and Robinson, M. and Rollins, J. G. and Romel, C. L. and Romie, J. H. and Ross, M. P. and Ryan, K. and Sadecki, T. and Sanchez, A. and Sanchez, E. J. and Sanchez, L. E. and Savage, R. L. and Schaetzl, D. and Schiworski, M. G. and Schnabel, R. and Schofield, R. M. S. and Schwartz, E. and Sellers, D. and Shaffer, T. and Short, R. W. and Sigg, D. and Slagmolen, B. J. J. and Soike, C. and Soni, S. and Srivastava, V. and Sun, L. and Tanner, D. B. and Thomas, M. and Thomas, P. and Thorne, K. A. and Todd, M. R. and Torrie, C. I. and Traylor, G. and Ubhi, A. S. and Vajente, G. and Vanosky, J. and Vecchio, A. and Veitch, P. J. and Vibhute, A. M. and von Reis, E. R. G. and Warner, J. and Weaver, B. and Weiss, R. and Whittle, C. and Willke, B. and Wipf, C. C. and Wright, J. L. and Yamamoto, H. and Zhang, L. and Zucker, M. E.",
  journal = {Phys. Rev. D},
  volume = {111},
  issue = {6},
  pages = {062002},
  numpages = {30},
  year = {2025},
  month = {Mar},
  publisher = {American Physical Society},
  doi = {10.1103/PhysRevD.111.062002},
}

@article{Virgo:2022ypn,
    author = "Acernese, F. and others",
    collaboration = "Virgo",
    title = "{The Virgo O3 run and the impact of the environment}",
    eprint = "2203.04014",
    archivePrefix = "arXiv",
    primaryClass = "gr-qc",
    doi = "10.1088/1361-6382/ac776a",
    journal = "Class. Quant. Grav.",
    volume = "39",
    number = "23",
    pages = "235009",
    year = "2022"
}

@article{Virgo:2022ysc,
    author = "Acernese, F. and others",
    collaboration = "Virgo",
    title = "{Virgo detector characterization and data quality: results from the O3 run}",
    eprint = "2210.15633",
    archivePrefix = "arXiv",
    primaryClass = "gr-qc",
    doi = "10.1088/1361-6382/acd92d",
    journal = "Class. Quant. Grav.",
    volume = "40",
    number = "18",
    pages = "185006",
    year = "2023"
}

@article{Janssens:2022xmo,
    author = "Janssens, Kamiel and Boileau, Guillaume and Christensen, Nelson and Badaracco, Francesca and van Remortel, Nick",
    title = "{Impact of correlated seismic and correlated Newtonian noise on the Einstein Telescope}",
    eprint = "2206.06809",
    archivePrefix = "arXiv",
    primaryClass = "astro-ph.IM",
    doi = "10.1103/PhysRevD.106.042008",
    journal = "Phys. Rev. D",
    volume = "106",
    number = "4",
    pages = "042008",
    year = "2022"
}

@article{LIGOScientific:2014pky,
    author = "Aasi, J. and others",
    collaboration = "LIGO Scientific",
    title = "{Advanced LIGO}",
    eprint = "1411.4547",
    archivePrefix = "arXiv",
    primaryClass = "gr-qc",
    doi = "10.1088/0264-9381/32/7/074001",
    journal = "Class. Quant. Grav.",
    volume = "32",
    pages = "074001",
    year = "2015"
}

@techreport{ETtaskforce:2025,
    title       = {The {ET} {B}aseline {D}etector {L}ayout},
    year        = {2025},
    author      = {{ETO Design Task Force}},
    number      = {EDMS: 3373921},
    institution = {ET collaboration},
    note         = {https://edms.cern.ch/document/3373921}
}

@article{Iacovelli:2026ixl,
    author = "Iacovelli, Francesco and Reali, Luca and Berti, Emanuele and Corsi, Alessandra and Sathyaprakash, B. S. and Wadekar, Digvijay",
    title = "{Not too close! Evaluating the impact of the baseline on the localization of binary black holes by next-generation gravitational-wave detectors}",
    eprint = "2604.11871",
    archivePrefix = "arXiv",
    primaryClass = "gr-qc",
    reportNumber = "ET-0169A-26",
    month = "4",
    year = "2026"
}

@article{Planck:2015fie,
    author = "Ade, P. A. R. and others",
    collaboration = "Planck",
    title = "{Planck 2015 results. XIII. Cosmological parameters}",
    eprint = "1502.01589",
    archivePrefix = "arXiv",
    primaryClass = "astro-ph.CO",
    doi = "10.1051/0004-6361/201525830",
    journal = "Astron. Astrophys.",
    volume = "594",
    pages = "A13",
    year = "2016"
}

@article{Thrane:2018qnx,
    author = "Thrane, Eric and Talbot, Colm",
    title = "{An introduction to Bayesian inference in gravitational-wave astronomy: parameter estimation, model selection, and hierarchical models}",
    eprint = "1809.02293",
    archivePrefix = "arXiv",
    primaryClass = "astro-ph.IM",
    doi = "10.1017/pasa.2019.2",
    journal = "Publ. Astron. Soc. Austral.",
    volume = "36",
    pages = "e010",
    year = "2019",
}

@article{LIGOScientific:2025yae,
    author = "Abac, A. G. and others",
    collaboration = "LIGO Scientific, VIRGO, KAGRA",
    title = "{GWTC-4.0: Methods for Identifying and Characterizing Gravitational-wave Transients}",
    eprint = "2508.18081",
    archivePrefix = "arXiv",
    primaryClass = "gr-qc",
    reportNumber = "LIGO-P2400300",
    month = "8",
    year = "2025"
}

@article{Dupletsa:2022scg,
    author = "Dupletsa, Ulyana and Harms, Jan and Banerjee, Biswajit and Branchesi, Marica and Goncharov, Boris and Maselli, Andrea and Oliveira, Ana Carolina Silva and Ronchini, Samuele and Tissino, Jacopo",
    title = "{gwfish: A simulation software to evaluate parameter-estimation capabilities of gravitational-wave detector networks}",
    eprint = "2205.02499",
    archivePrefix = "arXiv",
    primaryClass = "gr-qc",
    doi = "10.1016/j.ascom.2022.100671",
    journal = "Astron. Comput.",
    volume = "42",
    pages = "100671",
    year = "2023"
}

@techreport{ETDesign:2020,
    title       = {The {ET} {D}esign {R}eport {U}pdate },
    year        = {2020},
    author      = {{ET Steering Committee}},
    number      = {ET-0007A-20},
    note         = {https://apps.et-gw.eu/tds/?content=3\&r=16984},
    institution = {ET collaboration}
}

@article{Amann:2022pyq,
    author = "Amann, Florian and Badaracco, Francesca and DeSalvo, Riccardo and Naticchioni, Luca and Paoli, Andrea and Paoli, Luca and Ruggi, Paolo and Selleri, Stefano",
    title = "{Tunnel Configurations and Seismic Isolation Optimization in Underground Gravitational Wave Detectors}",
    eprint = "2204.04131",
    archivePrefix = "arXiv",
    primaryClass = "astro-ph.IM",
    doi = "10.3390/app12178827",
    journal = "Appl. Sciences",
    volume = "12",
    number = "17",
    pages = "8827",
    year = "2022"
}

@article{Fiori:2020arj,
    author = "Fiori, Irene and others",
    title = "{The Hunt for Environmental Noise in Virgo during the Third Observing Run}",
    doi = "10.3390/galaxies8040082",
    journal = "Galaxies",
    volume = "8",
    number = "4",
    pages = "82",
    year = "2020"
}

@article{VIRGO:2026okv,
    author = "Acernese, F. and others",
    collaboration = "VIRGO",
    title = "{Advanced Virgo during the LIGO-Virgo-KAGRA fourth observing run}",
    eprint = "2607.26872",
    archivePrefix = "arXiv",
    primaryClass = "gr-qc",
    reportNumber = "VIR-0259F-26",
    month = "7",
    year = "2026"
}

@misc{LogbookBoschi:2019,
    author = "Boschi, Valerio and others",
    key = "C130J Flyover Tests",
    howpublished = "\url{https://logbook.virgo-gw.eu/virgo/?r=44268}",
    year = "2019",
    note = "{V}irgo logbook entry",

}

@misc{LogbookFiori:2019,
key = "tractor test, report",
    author = "Fiori, Irene and others",
    year = "2019",
  howpublished = "\url{https://logbook.virgo-gw.eu/virgo/?r=46055}",
    note = "{V}irgo logbook entry"
}

@article{vanBeveren:2023seq,
    author = "van Beveren, Vincent and Bader, Maria and van den Brand, Jo and Jan Bulten, Henk and Campman, Xander and Koley, Soumen and Linde, Frank",
    title = "{A study of deep neural networks for Newtonian noise subtraction at Terziet in Limburg{\textemdash}the Euregio Meuse-Rhine candidate site for Einstein Telescope}",
    doi = "10.1088/1361-6382/acf3c8",
    journal = "Class. Quant. Grav.",
    volume = "40",
    number = "20",
    pages = "205008",
    year = "2023"
}

@techreport{Kong1992,
  author       = {A. Kong},
  institution  = {University of Chicago, Department of Statistics},
  number       = {Technical Report 348},
  year         = {1992},
}

@article{LIGOScientific:2026wfs,
    collaboration = "LIGO Scientific, VIRGO, KAGRA",
    title = "{GWTC-5.0: Observations from the Second Part of the Fourth LIGO-Virgo-KAGRA Observing Run and Updates to the Gravitational-Wave Transient Catalog}",
    eprint = "2605.27225",
    author = "Abac, A. G. and others",
    archivePrefix = "arXiv",
    primaryClass = "gr-qc",
    reportNumber = "LIGO-P2600152",
    month = "5",
    year = "2026"
}

@dataset{negri_2026_20799710,
  author       = {Negri, Luca},
  title        = {},
  month        = jun,
  year         = 2026,
  publisher    = {Zenodo},
  version      = {1.0},
  doi          = {10.5281/zenodo.20799710},
  url          = {https://doi.org/10.5281/zenodo.20799710},
}

@article{Caprini:2018mtu,
    author = "Caprini, Chiara and Figueroa, Daniel G.",
    title = "{Cosmological backgrounds of gravitational waves}",
    eprint = "1801.04268",
    archivePrefix = "arXiv",
    primaryClass = "astro-ph.CO",
    doi = "10.1088/1361-6382/aac608",
    journal = "Class. Quant. Grav.",
    volume = "35",
    number = "16",
    pages = "163001",
    year = "2018"
}

\appendix

\section{Supplementary Material}

\section{Correlated unlocks for continuous Markov chains}

While the continuous Markov chain formalism discussed previously can well model the duty cycle for current-generation gravitational-wave observatories, it doesn't model correlated downtimes, which is particularly relevant for co-located detectors, such as the three V-shaped detectors of the ET-$\Delta$ design. Correlated downtime can be accounted for by including a parameter $q_c < q_u$ that characterizes the rate at which multiple instruments are transitioning from online to offline. The resulting rate matrix is thus
\begin{equation}
    Q_{ij} =
    \begin{cases}
      (i-N)q_d - i q_u              & \text{if }i=j \text{ and } i<2, \\
      (i-N)q_d - i q_u + (i-1)q_c   & \text{if }i=j \text{ and } i\ge 2, \\
      (N-i)q_d                      & \text{if }j-i = 1, \\
      i q_u                         & \text{if }i=1 \text{ and } j=0,\\
      i (q_u  - q_c)                & \text{if }j-i = -1 \text{ and } i\ge 2,\\
      q_c                           & \text{if } i\ge 2 \text{ and } j=0, \\
      0                             & \text{else}.
    \end{cases}
\end{equation}
The associated stationary distributions for two- and three-detector networks are
\begin{equation}
  \pi(N=2) =  \frac{1}{\kappa_2} \begin{bmatrix}
  \alpha \gamma - \gamma + 2\\
  - 2 \alpha \left(\gamma - 2\right)\\
  2 \alpha^{2}\\
  \end{bmatrix}
\end{equation}
and
\begin{equation}
  \pi(N=3) =  \frac{1}{\kappa_3} \begin{bmatrix}
  2 \alpha^{2} \gamma - 4 \alpha \gamma^{2} + 7 \alpha \gamma + 2 \gamma^{2} - 7 \gamma + 6\\
  3 \alpha \left(\alpha \gamma + 2 \gamma^{2} - 7 \gamma + 6\right)\\
  - 6 \alpha^{2} \left(2 \gamma - 3\right)\\
  6 \alpha^{3}\\
  \end{bmatrix},
\end{equation}
respectively, where $\kappa_i = \prod^{N}_{i=1} \left[ i + i \alpha - (i-1)\gamma \right]$, $\alpha=q_d/q_u$ and $\gamma=q_c/q_u$.\\

\section{Testing the duty cycle model}
\begin{figure}[htp]
    \centering
    \includegraphics[width=\columnwidth]{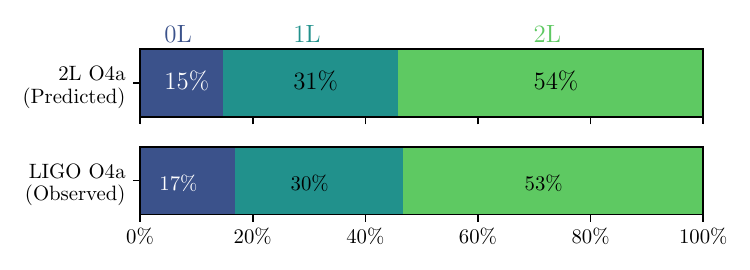}
    \caption{\red{Comparison of the predicted duty cycle from our model, considering 70\% of uptime of interferometers, with 10\% of correlated maintenance and the rest of the downtime due to uncorrelated causes. Our model is able to predict up to a few percentage points LIGO O4a data. The discrepancies can be attributed to correlated unlocks of non-co-located detectors, like the ones due to strong earthquakes, which are not considered in our model.} }
    \label{fig:O4a_comparison}
\end{figure}

\red{We compare in Fig. \ref{fig:O4a_comparison} the predicted duty cycle from our model for 2 non colocated interferometers to the O4a LIGO duty cycle. We select 70\% as the single interferometer uptime percentage, compatible with the mean between Hanford and Livingston in O4a, and allocate 10\% of the time to planned unlocks for maintenance and minor upgrades activities, as we use in this manuscript. The rest of the downtime is assigned to random uncorrelated events between detectors. Considering only the data used to fit the Markov chain model, the single detector uptimes were around 77\%, higher than the 70\% averge over the whole of O4a. Our model is able to be extrapolated to the complete O4a run and predict the uptimes for specific detector configurations with errors around 1\%. This difference can be attributed to correlated unlocks between non-colocated detectors that are not taken into account by the model, like strong distant earthquakes. The observed percentage of downtime due to earthquakes is around 2-4\% \cite{VIRGO:2026okv}, which would be able to explain the discrepancy between the model and the observation. This result supports the validity of our model and its assumptions.} 

\section{Fisher information matrix and full sampling}
\label{app:fisher_matrix_vs_PE}
\begin{figure}[htp]
    \centering
    \includegraphics[width=\columnwidth]{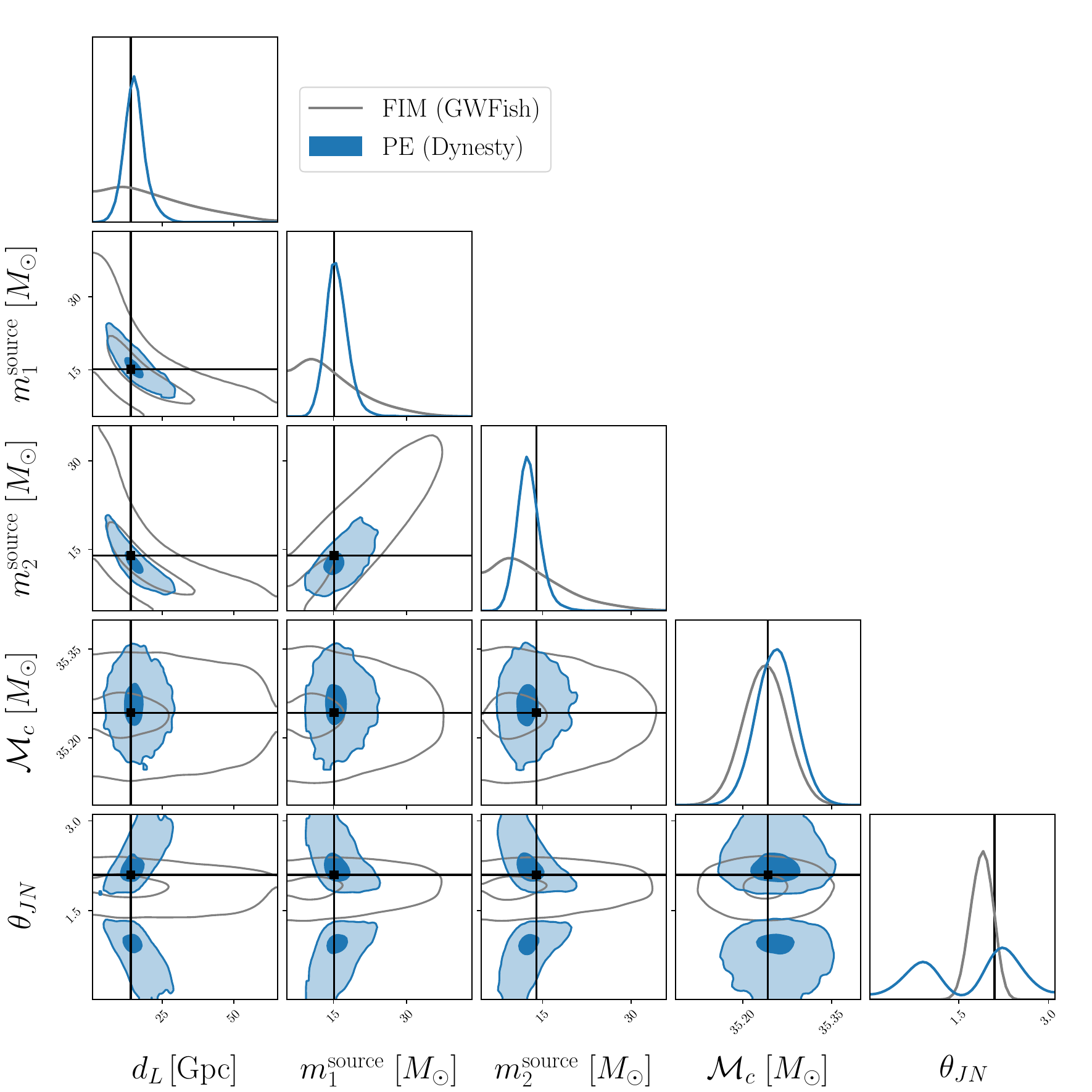}
    \caption{Comparison of results between GWfish and dynesty on the same injection for an ET 2V network and SNR of 28. While the width of the chirp mass posterior aligns with the nested sampling estimate, FIM fails to capture the complex correlation structure between the inclination angle and the luminosity distance, which leads to an overall overestimation of the latter.}
    \label{fig: Fisher vs dynesty}
\end{figure}

The comparison between the  2 proposed designs of the Einstein Telescope has been the subject of many different works~\cite{Branchesi:2023mws}, the majority of which rely on the \ac{FIM} formalism. This tool can give quick estimates of the width of the credible intervals for simulated events. Classic parameter estimation algorithms, such as nested sampling, which are routinely used to perform parameter estimation on current GW events~\cite{LIGOScientific:2025yae}, require tens to hundreds of CPU hours to give such estimates, making these methods computationally unfeasible to analyze a whole catalogue of ET injections spanning a year of observations. The FIM formalism is, however, known to give imprecise estimates in many instances~\cite{Vallisneri:2007ev}: estimating the whole posterior based on the likelihood morphology around a single point will lead to large inaccuracies if the posteriors distribution has a complex correlation structure or presents multimodalities. This is often the case for gravitational wave posteriors, with the morphology becoming more complex the fewer detectors are present in the network. We thus cannot rely on the FIM formalism to test the capabilities of the small detector networks, which are the focus of this work. In Fig. \ref{fig: Fisher vs dynesty}, we compare a posterior obtained with the FIM formalism with the package GWfish ~\cite{Dupletsa:2022scg}, implementing prior effects with the procedure highlighted in \cite{Dupletsa:2024gfl}, and nested sampling estimates for an event with SNR of 28 as observed by ET-2V. Parameters such as the chirp mass are well predicted, whereas FIM severely overestimates the distance posterior and consequently the source frame masses. FIM also gives a unimodal inclination angle ($\theta_{JN}$) while nested sampling recovers the two modes. This is a consequence of the FIM approximating the posterior as a unimodal Gaussian, which artificially removes existing degeneracies. We find that, on average, \ac{FIM} overrestimates the luminosity distance credible intervals by a factor 2. We also find this overshoot to be more prevalent for the full ET-$\Delta$, where this overshoot factor is around 2.6, while it is less prevalent for the full ET-2L, where this factor is on average 1.8  

\subsection{More detailed comparison between detector configurations}

Here, we expand the comparison between credible intervals as obtained by analyzing the data with different network combinations.\\

Comparing credible intervals obtained for source frame masses between the full ET-2L and the full ET-$\Delta$, we find the median ratio to be $\sigma_{\mathrm{2L}}/\sigma_\Delta =  0.9$, implying that the ET-2L posteriors are 10\% narrower than ET-$\Delta$. This result roughly aligns with previous works, highlighting how a network of 2 misaligned L detectors has access to both GW polarizations like ET-$\Delta$, and the higher SNRs made possible by the longer arm lengths allow for tighter constraints.  When comparing the detector networks where only one detector is online (ET-1V and ET-1L) we find that  $\sigma_{\mathrm{1L}}/\sigma_{\mathrm{1V}} = 0.93$, highlighting the fact that, when detector networks have comparable geometries, their performance is mainly driven by the \ac{SNR} \\ 


When detector networks lose interferometers, we expect the average width of confidence intervals to increase. As highlighted before, when one detector of ET-$\Delta$ is offline, there is no qualitative information being lost. 
We find that for source masses the median credible interval ratio between the full ET-$\Delta$ and ET-$\Delta$ with one detector offline is $\sigma_{\Delta}/\sigma_{\mathrm{2V}} = 0.88$, while when $2$  \red{detector}s are offline, the average $\sigma_{\Delta}/\sigma_{\mathrm{1V}} = 0.49$, meaning that credible intervals are almost twice as wide as the full configuration. This reflects the fact that one  \red{detector} cannot localize the source well, leading to large errors in luminosity distance, which, in turn, leads to large errors for source masses.\\

Regarding the loss of one interferometer in the ET-2L, there is a considerable widening of the recovered credible intervals. For source masses, we find the average $\sigma_{\mathrm{1L}}/\sigma_{\mathrm{2L}} = 0.47$. Like in the $\sigma_{\Delta}/\sigma_{\mathrm{1V}}$, the credible intervals are now more than twice as wide.

So far, we have focused on parameters for which the recovery is linked to how well the geometry of the source can be reconstructed. When looking at other parameters that are not strictly dependent on detector geometry, like spins, the widening of credible intervals is mainly driven by the SNR of the different configurations. Taking $\chi_{\rm{eff}}$ as an example, we find that $\sigma_{\mathrm{2L}}/\sigma_{\mathrm{1L}} = 0.74$, roughly in line with the SNR ratio between these two configurations, likewise $\sigma_{\mathrm{2V}}/\sigma_{\mathrm{1L}} = 1.12$, since ET-2V has on average lower SNR.\\

\section{Additional cornerplots}
\begin{figure*}[htp]
    \centering
     \includegraphics[width=\textwidth]{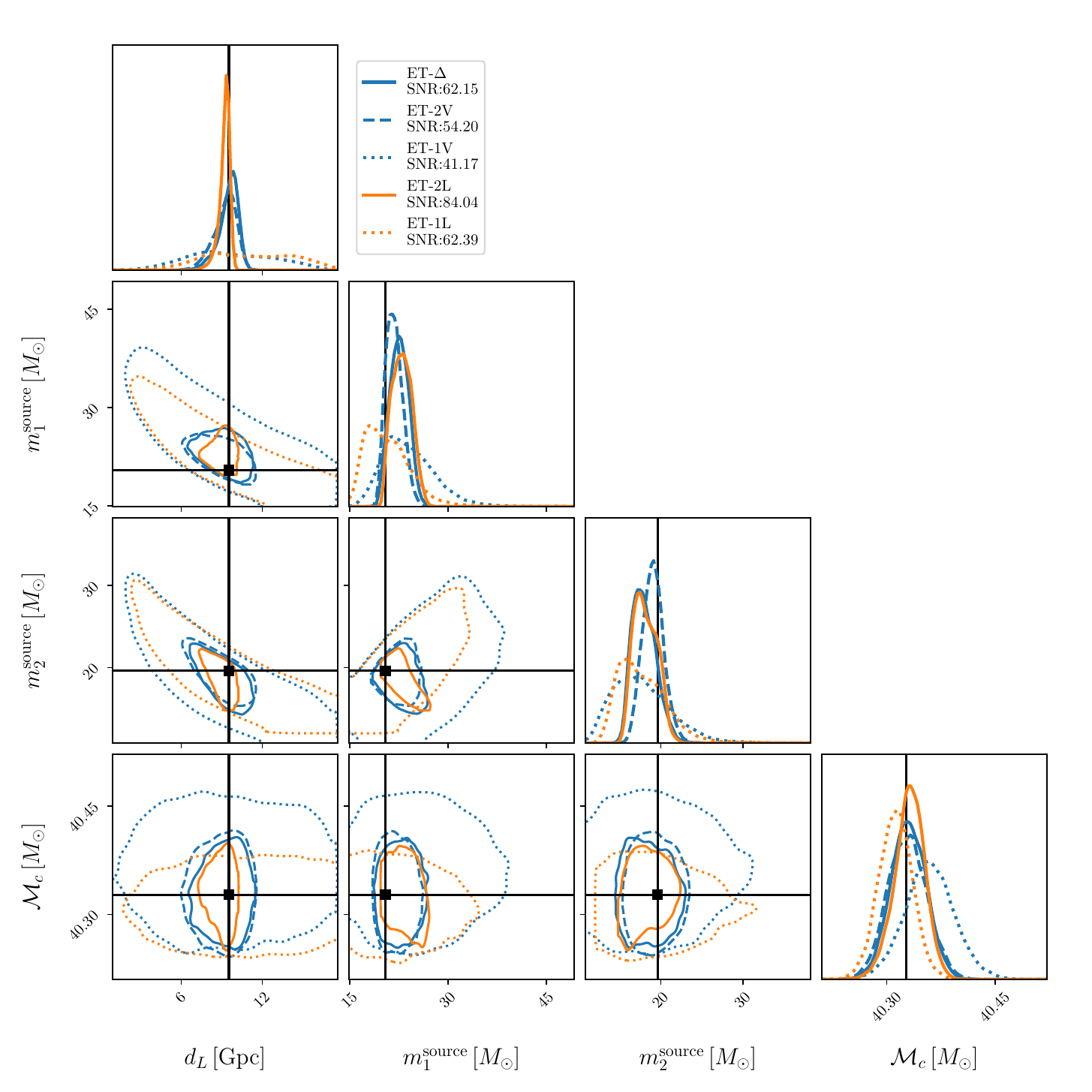}
    
    \caption{All cornerplots overlayed for the 5 different configurations for the signal in Figure~\ref{fig: example single injection 2V vs 1L}. It is evident how the ET-2L posteriors widen as soon as 1  \red{detector} is lost, for source masses and distance, while for ET-$\Delta$, the posteriors with one detector offline are similar to the full configuration.}
    \label{fig: all configuration injection 19}
\end{figure*}

\begin{figure*}[htp]
    \centering
    \includegraphics[width=\textwidth]{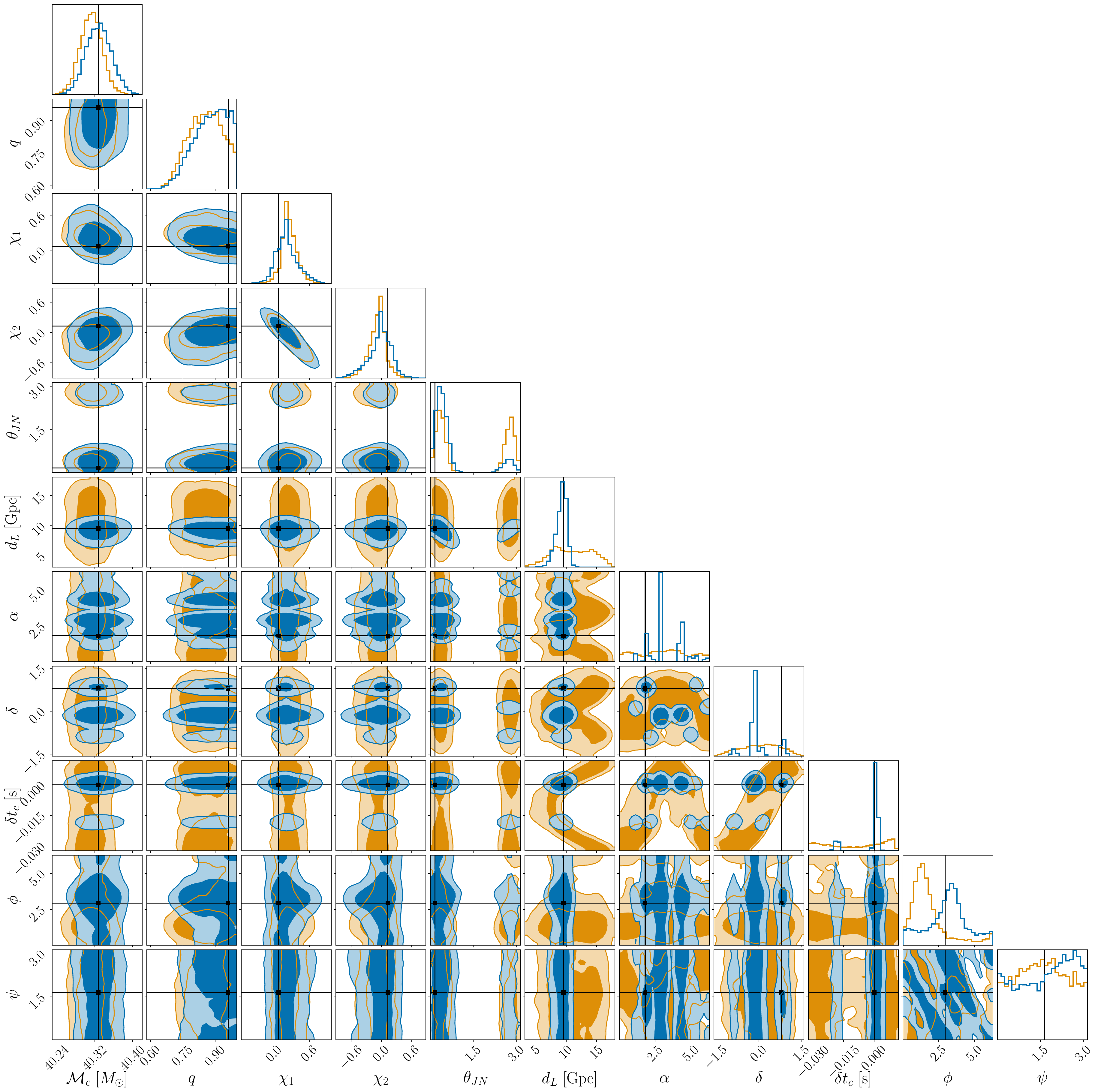}
    \caption{Full cornerplot for the signal in Figure~\ref{fig: example single injection 2V vs 1L}. While some intrinsic parameters are better constrained by ET-1L due to the higher SNR, other extrinsic parameters are better constrained by the ET-$2$V due to the better measurement of the polarization of the \ac{GW} signal.}
    \label{fig: example single injection 2V vs 1L full cornerplot}
\end{figure*}

Fig. \ref{fig: all configuration injection 19} shows a corner plot of all of the five studied configurations for the injection presented already in Fig. \ref{fig: example single injection 2V vs 1L}. It is evident how the posterior of the full ET-2L design is narrower than the full ET-$\Delta$ one. However, we can notice that ET-2L with one detector offline (ET-1L) is much wider than ET-$\Delta$ with one detector offline (ET-2V) and actually more similar to ET-$\Delta$ with 2 detectors offline (ET-1V). This plot also shows how ET-2V is very close to the posterior of ET-$\Delta$.\\
Fig. \ref{fig: example single injection 2V vs 1L full cornerplot} shows the full sampled parameter space for the same injection shown in fig. \ref{fig: example single injection 2V vs 1L}.  Parameters like the detector frame chirp mass are better constrained by ET-1L due to its higher SNR, while extrinsic parameters like luminosity distance are better constrained by ET-2V. 

\end{document}